\documentclass[aps,prc,twocolumn,superscriptaddress,nofootinbib,preprintnumbers]{revtex4-2}

\usepackage{xcolor}
\usepackage{amsmath}
\usepackage{amssymb}
\usepackage{graphicx}
\usepackage{multirow}
\begin{document}
\preprint{}
%\preprint{}

\title{Bulk viscosity from neutron decays to dark baryons in neutron star matter}


\author{Steven P.~Harris}
%\email[]
\affiliation{Department of Physics and Astronomy, Iowa State University, Ames, IA, 50011, USA}
\affiliation{Center for the Exploration of Energy and Matter and Department of Physics,
Indiana University, Bloomington, IN 47405, USA}
\author{C.J.~Horowitz}
%\email[]
\affiliation{Center for the Exploration of Energy and Matter and Department of Physics,
Indiana University, Bloomington, IN 47405, USA}

\date{April 16, 2026}
\begin{abstract}
The existence of a dark baryon that mixes with the neutron leads to the possibility of neutron decay into a dark sector.  Such dark decays have been studied as possibly relevant for the neutron decay anomaly and for their potential impacts on neutron stars.  The most popular formulation is a dark sector consisting of a dark baryon $\chi$ and a dark scalar $\phi$, where a neutron in vacuum decays 1\% or less of the time via the channel $n\rightarrow \chi+\phi$.  In this work, we consider the effect of this additional neutron decay channel on transport in neutrons star mergers.  We find that the neutron dark decay rate in medium is quite slow, and thus the dark baryons modify the dense matter equation of state in a way that decreases the Urca bulk viscosity by, at most, a factor of 2-3.  However, if the neutron dark decay was to occur more rapidly, then the bulk viscosity at merger temperatures of tens of MeV would be strongly enhanced, potentially rapidly damping oscillations in merger environments and therefore providing a signature of slowly equilibrating matter in the merger. 
\end{abstract}
\maketitle
%%%%%%%%%%%%%%%%%%%%%%%%%%%%%%%%%%%%%%%%%%%%%%%%%%%%%%%%%%%%%%%%%%%%%%%%%%%%%
\section{Introduction}

As matter is compressed from terrestrial conditions to increasingly neutron-rich nuclei and beyond, neutrons become increasingly prevalent, culminating in an upwards of 90\% abundance in the uniform neutron-proton-electron ($npe$) Fermi liquid that makes up the matter in a neutron star \cite{Schaffner-Bielich:2020psc,Chatziioannou:2024jsr}.  Thus, the properties of the neutron critically influence the behavior of astrophysical systems containing dense matter.  Vacuum properties of the neutron, including its lifetime and various decay channels, can impact the kilonova that occurs in the wake of a neutron star merger \cite{Metzger:2019zeh,Metzger:2014yda}.  In-medium, the neutron dispersion relation is modified \cite{Glendenning:1997wn} and decay channels that may be kinematically blocked in vacuum become possible.  Neutron reactions in medium play a vital role in both neutron star cooling and bulk-viscous dissipation \cite{Yakovlev:2000jp,Schmitt:2017efp,Harris:2024evy}

Over the past couple decades, the neutron lifetime has been measured with increasing precision.  Two measurement techniques, the beam method and the bottle method, exist but their results are in tension with one another \cite{Wietfeldt:2011suo,Wietfeldt:2018upi,Dubbers:2011ns,Dubbers:2021wqv}.  In the beam method, a beam of cold neutrons passes through a chamber where protons resulting from neutron decays are extracted from the beam and counted.  The bottle method involves ultracold neutrons that are contained and the surviving neutrons are periodically counted.   At present, the ``beam method'' lifetime is $\tau_{\text{beam}}=888.0\pm 2.0$ s, while the ``bottle method'' lifetime is $\tau_{\text{bottle}}=878.4\pm0.5$ s \cite{Fornal:2023wji}.  This tension, now over $4\sigma$ \cite{Wietfeldt:2018upi}, is termed the ``neutron decay anomaly'' (or ``neutron lifetime anomaly'').

In 2018, Fornal \& Grinstein proposed that this discrepancy could be due to new physics, in particular, the existence of a dark sector into which the neutron decays about 1\% of the time \cite{Fornal:2018eol}.  In this formulation, the bottle method measures the correct neutron lifetime, while the beam method only measures the neutron decay rate to protons, missing out on an additional dark decay channel.  Many possible dark sectors into which the neutron may decay have been proposed \cite{Fornal:2018eol,Cline:2018ami,Ivanov:2018vit,Strumia:2021ybk,Elahi:2020urr,Berezhiani:2018udo}, including
\begin{subequations}
\begin{align}
    n&\rightarrow \chi+\gamma,\label{eq:solution1}\\
    n&\rightarrow \chi+ e^+ + e^-,\label{eq:solution2}\\
    n&\rightarrow \chi + \phi,\label{eq:solution3}\\
    n&\rightarrow \chi + A',\label{eq:solution5}\\
    n&\rightarrow \tilde{\chi}+\tilde{\chi}+\tilde{\chi},\label{eq:solution4}\\
    n&\rightarrow \chi + \nu+\bar{\nu},\label{eq:solution6}
\end{align}
\end{subequations}
where $\gamma$, $e^{\pm}$, and $\nu/\bar{\nu}$ are the usual photon, electron/positron, and neutrino/antineutrino, $\chi$ is a dark baryon, $\phi$ is a dark scalar (or pseudoscalar), $\tilde{\chi}$ is a dark quark (with baryon number 1/3), and $A'$ is a dark gauge boson (abelian or nonabelian).  These reactions can be generated by considering mass mixing between the neutron and a dark baryon, for example, $n\rightarrow \chi+\gamma$ can be generated through the mixing and the neutron magnetic moment interaction \cite{Fornal:2018eol}, though one can also consider a particular reaction independent of an explicit mixing model.

Out of the processes \ref{eq:solution1}-\ref{eq:solution6}, the two processes involving photons and electron/positron pairs were quickly experimentally ruled out \cite{Tang:2018eln, UCNA:2018hup}, but solutions where the neutron decays entirely into dark particles (\ref{eq:solution3}-\ref{eq:solution4}) are more difficult to completely exclude.  The vast majority of work has focused on the dark baryon $\chi$ \cite{Fornal:2018eol,McKeen:2018xwc,Motta:2018bil,Motta:2018rxp,Baym:2018ljz,Grinstein:2018ptl,Gresham:2018rqo,Husain:2022bxl,Husain:2023fwb,Shirke:2023ktu,Routaray:2024lni,Bastero-Gil:2024kjo,Das:2025pjl,Shirke:2024ymc,Divaris:2025ftf}, and some of it specifically on reaction \ref{eq:solution3} \cite{Husain:2022bxl,Husain:2023fwb,Shirke:2023ktu}, where the neutron decays into $\chi$ and a dark scalar $\phi$.  This is the solution we examine in this work.  

Along the same lines, solutions to the neutron decay anomaly involving mirror dark sectors have been proposed.  In mirror models, a ``dark'' copy of the standard model exists, and the light and dark sectors are coupled by some cross interaction \cite{Berezhiani:2018udo,Berezhiani:2018eds,Berezhiani:2020zck,Tan:2023mpj,Okun:2006eb,Berezhiani:2005ek}.  The degeneracy between the dark and light sectors allows neutral light particles like the neutron \cite{Berezhiani:2005hv} to oscillate into their dark counterparts.  There may or may not be small mass differences between the two.  These oscillations have been proposed as solutions to the neutron decay anomaly as well \cite{Berezhiani:2018udo,Berezhiani:2018eds,Tan:2023mpj}.

In 2018, Fornal \& Grinstein \cite{Fornal:2018eol} proposed that the neutron dark decay branching ratio would need to be about 1\% in order to resolve the difference between the beam and bottle neutron lifetime measurements.  Soon after, Czarnecki, Marciano, \& Sirlin \cite{Czarnecki:2018okw} pointed out that the measured value of $g_A$ has increased over time, from about 1.2 to roughly 1.27.  If one discards older (pre-2002) measurements of $g_A$ and considers only the newer measurements of $g_A$, then the theoretical standard-model calculation of the neutron lifetime agrees well with the bottle measurement (see, e.g., \cite{Czarnecki:2018okw,Belfatto:2019swo,Dubbers:2021wqv} for the precise numbers), leaving no room for a dark branching ratio of 1\%.  This analysis was updated by Dubbers \textit{et al.}~\cite{Dubbers:2018kgh}.  Czarnecki, Marciano, \& Sirlin \cite{Czarnecki:2018okw} put a 95\% confidence level upper bound on the dark branching ratio of 0.27\%, and the reanalysis by Dubbers \textit{et al.}~\cite{Dubbers:2018kgh} updates that bound to 0.15\%, if the pre-2002 $g_A$ measurements are removed from the average.  However, Gardner \& Zakeri \cite{Gardner:2023wyl} note that recent measurements of correlation coefficients complicate the picture presented in \cite{Dubbers:2018kgh,Czarnecki:2018okw}.  Importantly, none of these works dispute that neutron dark decays are allowed, even if they are unable to produce the 1\% branching ratio needed to solve the lifetime discrepancy.  

After the initial solutions to the lifetime anomaly \cite{Fornal:2018eol,Cline:2018ami,Ivanov:2018vit,Berezhiani:2018udo} were proposed, researchers turned to studying the consequences of dark baryons on neutron stars.  Immediately, one is confronted with the question of whether the neutron star consists of one fluid or two.  In many studies, it is assumed that the $\chi$ interacts strongly enough with the nuclear matter that the neutron star consists of one fluid, and one can consider an $npe\chi$ equation of state (EoS) \cite{Baym:2018ljz,McKeen:2018xwc,Cline:2018ami,Motta:2018bil,Motta:2018rxp,Grinstein:2018ptl,Husain:2022bxl,Husain:2023fwb,Shirke:2023ktu,Routaray:2024lni,Shirke:2024ymc,Bastero-Gil:2024kjo,Divaris:2025ftf}.  Typically, the $\phi$ is assumed to be light and thus it escapes from a neutron star (barring an absorption process, which is not usually considered).  It was determined that the introduction of $\chi$ particles as a free Fermi gas would destabilize neutron stars due to their softening of the EoS \cite{Baym:2018ljz,McKeen:2018xwc,Motta:2018rxp,Motta:2018bil}.  To counteract this, the dark baryons are given a repulsive self-interaction which stiffens the EoS and allows for neutron stars to have masses in excess of two solar masses \cite{McKeen:2018xwc,Cline:2018ami,Motta:2018bil,Grinstein:2018ptl,Husain:2022bxl,Husain:2023fwb,Shirke:2023ktu,Routaray:2024lni,Shirke:2024ymc}.  The impact on the EoS of dense matter due to a population of chemically equilibrated dark baryons is a topic of continued study.

If, on the other hand, the dark baryons interact only very weakly with the nuclear matter, but reasonably strongly with each other, then the dark matter will form a second fluid in the neutron star, and the two fluids will interact predominantly gravitationally.  This case was examined by \cite{Ellis:2018bkr} and, in the case of mirror matter, by \cite{Berezhiani:2020zck}.  Finally, the dark baryons produced in the neutron star might not be prevalent enough to form their own fluid, but can still have effects on neutron stars \cite{Berryman:2022zic,Alonso-Alvarez:2021oaj,Berryman:2023rmh,Gardner:2023wyl}.  Aside from the motivation of solving the neutron decay anomaly, many studies of one-fluid \cite{Panotopoulos:2017idn,Das:2018frc,Issifu:2024htq,Thakur:2024btu,Lenzi:2022ypb,Lourenco:2022fmf,Das:2018frc,Lopes:2024ixl,Shahrbaf:2024gdm,Shahrbaf:2025hsw} or two-fluid \cite{Gresham:2018rqo,Ellis:2018bkr,Liu:2024rix,Caballero:2024qtv,Kumar:2025yei,Thakur:2023aqm,Avila:2023rzj,Giangrandi:2025rko,Giangrandi:2022wht,Koehn:2024gal,Ellis:2017jgp,Emma:2022xjs,Sotani:2025hzb,Kain:2021hpk,Gleason:2022eeg,Cronin:2023xzc,Hajkarim:2024ecp,Shawqi:2025cca,Issifu:2025gsq,Leung:2011zz,Rutherford:2022xeb,Shakeri:2022dwg,Miao:2022rqj,Karkevandi:2021ygv,Berezhiani:2020zck,Goldman:2019dbq} dark-matter-admixed neutron stars have been done, where dark matter has been found to modify the mass-radius relation, oscillation spectra, and cooling of isolated neutron stars and the dynamics of supernovae \cite{Issifu:2024htq,Fiorillo:2024upk,Alonso-Alvarez:2021oaj} and neutron star mergers \cite{Giangrandi:2025rko,Koehn:2024gal,Ellis:2017jgp,Emma:2022xjs}.

Putting aside dark matter for a moment, the study of beta equilibration and bulk viscosity in neutron stars has progressed rapidly over the past decade or two.  Beta equilibration occurs in $npe$ matter through the Urca process: neutron decay and electron capture \cite{Yakovlev:2000jp}.  Oscillations in the neutron star push the matter out of beta equilibrium, leading the Urca process to change the proton fraction to attempt to restore equilibrium.  These out-of-equilibrium processes give rise to bulk-viscous dissipation, which damps density oscillations \cite{Harris:2024evy,Schmitt:2017efp}.  Calculations of the bulk viscosity of $npe$ matter in cold neutron stars, where the matter is very slow to beta equilibrate, began many decades ago \cite{1968ApJ...153..835F} and have improved since \cite{Sawyer:1989dp,Haensel:1992zz,Haensel:2000vz,Haensel:2001mw,Gusakov:2007px,Alford:2010gw}, and the results have been applied to studying the bulk-viscous damping of neutron star oscillations \cite{1990ApJ...363..603C,Gusakov:2005dz} including r-modes \cite{Andersson:2024amk,Andersson:1998ze,Alford:2010fd,Alford:2011pi}.  Bulk viscosity in other phases of matter has been studied (for reviews, see \cite{Harris:2024evy,Schmitt:2017efp}), including matter with multiple equilibration channels \cite{Sad:2007afd,Alford:2006gy}, which leads to a more complicated bulk viscosity.

In recent years, with the measurement of the gravitational wave signal from the inspiral of the neutron star merger GW170817 \cite{LIGOScientific:2017vwq}, the hot, dense matter in a neutron star merger remnant became a major focus of study.  This matter reaches temperatures of many tens of MeV \cite{Perego:2019adq,Hammond:2021vtv,Fields:2023bhs,particles2010004}, hot enough to significantly shorten the neutrino mean free path, and fluid elements within the remnant experience dramatic density oscillations as the stars collide and merge into one object \cite{Alford:2017rxf}.  Alford \textit{et al.}~\cite{Alford:2017rxf} did an estimate of the bulk viscosity of $npe$ matter in neutron star merger conditions and found that the bulk viscosity could damp density oscillations on timescales relevant to neutron star mergers.  Follow-up calculations showed that if the $npe$ matter is transparent to neutrinos (which is true for $T\lesssim 5\text{ MeV}$), bulk viscous dissipation from the Urca process can damp density oscillations in as little as 5 ms \cite{Alford:2019qtm,Alford:2023gxq}, while if the neutrino mean free path is short enough for the neutrinos to be thermally equilibrated with the $npe$ matter, beta equilibration is too fast to yield significant bulk viscosity \cite{Alford:2019kdw}.  These conclusions persisted, even as new degrees of freedom like muons \cite{Alford:2021lpp,Alford:2022ufz,Alford:2023uih} and pions \cite{Harris:2024ssp} and their equilibration channels were considered.  Recent calculations of the bulk viscosity in dense quark matter show that its bulk viscosity is highest at relatively low temperatures as well \cite{CruzRojas:2024etx,Hernandez:2024rxi,Alford:2024tyj,Alford:2025tbp}, and may not be too different from that of $npe$ matter in some regimes \cite{Alford:2024tyj,Alford:2025tbp}.

Because bulk viscosity does appear to be relevant to neutron star mergers when the temperature is around a few MeV, but this temperature is also where neutrinos begin to be trapped, suppressing the bulk viscosity, implementing the Urca reactions in numerical simulations of neutron star mergers in the context of a neutrino-transport scheme \cite{Foucart:2022bth} is important.  To date, several merger simulations have studied beta equilibration effects \cite{Most:2022yhe,Espino:2023dei,Zappa:2022rpd}, and some have seen modifications of the postmerger gravitational wave signal, but others have not.  Another approach is to pre-calculate the bulk viscosity and then implement it in a Muller-Israel-Stewart scheme in the simulation \cite{Chabanov:2023abq,Chabanov:2023blf,Camelio:2022fds,Camelio:2022ljs}.  In any case, one major factor acting against large bulk-viscous dissipation is that much of the matter in the remnant (especially after the first few milliseconds) is at temperatures of several tens of MeV, well above where the Urca bulk viscosity is large \cite{Espino:2023dei,Most:2021zvc}.

There has been increased interest in recent years as to how dark matter or beyond the standard model (BSM) particles can enhance transport processes in compact objects.  Feebly-interacting particles produced within a neutron star or supernovae core will escape from the object, if they are sufficiently light, cooling it \cite{Raffelt:1996wa,Caputo:2024oqc,Dietrich:2019shr,Harris:2020qim,Beznogov:2018fda,Alonso-Alvarez:2021oaj} and perhaps generating additional electromagnetic signatures if the particle is unstable \cite{Dev:2023hax,Diamond:2023cto,Muller:2023vjm,Diamond:2023scc,Diamond:2021ekg,Jaeckel:2017tud}.  If the particles are slightly more strongly interacting, they will be marginally trapped in the system and can contribute, for example, to energy transfer between the core and mantle of a supernovae \cite{Caputo:2022mah,Fiorillo:2025yzf}.  In the diffusive regime, this transport can be described in terms of an enhanced thermal conductivity or shear viscosity \cite{Horowitz:2012jd,Dev:2021kje}.

Chemical reactions of dark sector particles in dense matter environments, typically motivated by the neutron decay anomaly, have been studied only in recent years.  Berryman \textit{et al.}~\cite{Berryman:2022zic} consider baryon-number-violating processes in neutron stars; a dark decay like $n\rightarrow\chi+\phi$ is an example of a process that \textit{appears} to violate baryon number as the baryon number of the neutron is converted to the dark baryon.  They consider the possibility that a neutron will decay into a dark baryon, which then will leave the $npe$ matter out of beta equilibrium, triggering Urca reequilibration.  This phenomenon of two possible beta equilibration channels involving the neutron was taken up by Shirke \textit{et al.}~in a study of the effect of neutron dark decays on the r-mode instability window \cite{Shirke:2023ktu}.  They found that a rapid neutron decay into a dark sector enhances the bulk viscosity and thus shrinks the r-mode instability window, however they did not perform a specific calculation of the neutron dark decay rate itself and their calculation of the bulk viscosity of this multi-channel system was flawed.  

In this work, we consider the $n\rightarrow\chi+\phi$ solution to the neutron decay anomaly and assume that the $\chi$ is thermally equilibrated with the $npe$ matter, while the $\phi$ escapes.  We do a full calculation of both decay rates of the neutron (Urca and the dark decay $n\rightarrow\chi+\phi$) and then calculate the bulk viscosity, taking into account the coupled nature of these decays (as illuminated by \cite{Berryman:2022zic}).  In Sec.~\ref{sec:n_decay_anomaly}, we describe the solution to the neutron lifetime anomaly where the neutron decays into a dark baryon $\chi$ and a dark scalar $\phi$ and in Sec.~\ref{sec:npechi_matter} we describe the EoS of dense matter that contains thermally-equilibrated $\chi$ particles as well as the standard $npe$ matter.  This includes a discussion of the self-repulsion of the dark baryons.  In Sec.~\ref{sec:bulkviscosity} we derive the bulk viscosity in a system where the neutron can decay through two different channels, arriving at a tractable formula for the bulk viscosity.  Finally, in Sec.~\ref{sec:results}, we calculate the rates of the two neutron decays, which includes the use of the novel nucleon width approximation (NWA) for the neutron dark decay rate.  We calculate the bulk viscosity of this system and then also study the bulk viscosity as we increase the $n\chi\phi$ coupling beyond that which leads to a 1\% branching ratio (in vacuum), as this leads to an interesting general point about moderate-speed chemical reactions in the high-temperature conditions of neutron star mergers.  

We work in natural units, where $\hbar=c=k_B=1$.  All data presented in the figures can be found in the Zenodo repository \cite{harris_2025_17315839}.
%%%%%%%%%%%%%%%%%%%%%%%%%%%%%%%%%%%%%%%%%%%%%%%%%%%
\section{Neutron decay anomaly}\label{sec:n_decay_anomaly}
Fornal \& Grinstein proposed that the neutron decay anomaly can be explained if, in vacuum, 1\% of neutron decays are into a dark sector, while the other 99\% are the standard $n\rightarrow p+e^-+\bar{\nu}_e$ \cite{Fornal:2018eol,Fornal:2023wji}.  Even if the dark decay can no longer resolve the lifetime discrepancy, as has been suggested recently \cite{Czarnecki:2018okw,Dubbers:2018kgh,Berezhiani:2018udo,Belfatto:2019swo,Dubbers:2021wqv} (though see \cite{Gardner:2023wyl}), we will use the 1\% branching ratio as a benchmark in this work, and also discuss our results with the stricter branching ratio limit of 0.15\% \cite{Dubbers:2018kgh}.

While many different dark decay channels have been proposed (\ref{eq:solution1}-\ref{eq:solution6}), in this work we will consider just  
\begin{equation}
    n\rightarrow\chi+\phi,
\end{equation}
where $\chi$ is a dark baryon and $\phi$ is a dark scalar, with masses $m_{\chi}$ and $m_{\phi}$, respectively.  In future sections, we will study dense matter with the Lagrangian
\begin{equation}
    \mathcal{L} = \mathcal{L}_{npe}^{\text{free}}+\mathcal{L}_{npe}^{\text{int.}}+\mathcal{L}_{\chi\phi}^{\text{free}}+\mathcal{L}_{\chi}^{\text{int.}}+\mathcal{L}_{n\chi\phi}^{\text{int.}}.\label{eq:total_Lagrangian}
\end{equation}
Here we focus on just the $n\chi\phi$ interaction, which has the the Lagrangian
\begin{equation}
    \mathcal{L}_{n\chi\phi}^{\text{int.}} = g_{\phi}(\bar{\chi}n+\bar{n}\chi)\phi.
\end{equation}

Given this Lagrangian, we can calculate the neutron dark decay rate in vacuum.  The $n\chi\phi$ vertex factor is $ig_{\phi}$ and the spin-summed matrix element is
\begin{equation}
    \sum_{\text{spins}}\vert\mathcal{M}\vert^2 = 2g_{\phi}^2\left[\left(m_n+m_{\chi}\right)^2-m_{\phi}^2\right]. \label{eq:nchiphi_matrix_element}
\end{equation}
The resultant decay rate, which uses the matrix element averaged over the spin of the initial neutron (that is, Eq.~\ref{eq:nchiphi_matrix_element} divided by two), is
\begin{align}
    \Gamma_{n,\text{ dark}} &= \dfrac{g_{\phi}^2}{16\pi m_n^3}\left[\left(m_n+m_{\chi}\right)^2-m_{\phi}^2\right]^{3/2}\\
    &\times \left[\left(m_n-m_{\chi}\right)^2-m_{\phi}^2\right]^{1/2},  \nonumber
\end{align}
where it is also required that $m_n > m_{\chi}+m_{\phi}.$

\begin{figure}\centering
\includegraphics[width=0.45\textwidth]{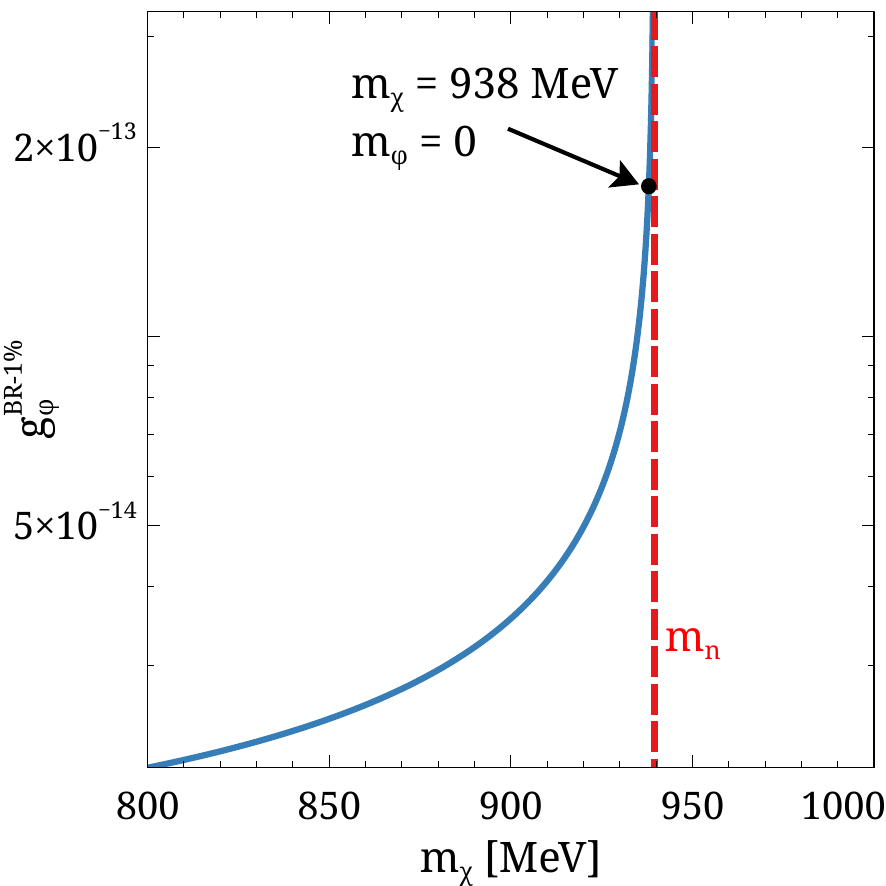}
\caption{The value of $g_{\phi}$ that yields a 1\% branching ratio, as a function of the mass of the $\chi$, with $m_{\phi}=0$.  The fiducial value chosen in this paper is represented by the black dot.}
\label{fig:gphiplot}
\end{figure}

Demanding that this rate match 1\% of the measured (by the bottle method) neutron decay rate yields a relationship between the three model parameters $\left\{g_{\phi},m_{\chi},m_{\phi}\right\}$.  It is most sensible to treat this relationship as a function for $g_{\phi}$, meaning that the dark matter model now has just two parameters $\left\{m_{\chi},m_{\phi}\right\}$ and $g_{\phi}$ is chosen to yield the 1\% branching ratio $\Gamma_{n,\text{ dark}} = 1/(878.4 \text{ s})/100 = 1.138\times 10^{-5} \text{ s}^{-1}$ \cite{Fornal:2023wji}, meaning
\begin{align}
    g_{\phi}^{\text{BR}-1\%} = \dfrac{1.767\times 10^{-8}}{\left[\left(m_n+m_{\chi}\right)^2-m_{\phi}^2\right]^{3/4}\left[\left(m_n-m_{\chi}\right)^2-m_{\phi}^2\right]^{1/4}},
\end{align}
where the masses are measured in MeV.  In Fig.~\ref{fig:gphiplot} we plot $g_{\phi} = g_{\phi}^{\text{BR}-1\%}(m_{\chi},m_{\phi})$ and see that a typical value\footnote{If we instead chose the coupling to be pseudoscalar, considering the $\phi$ to be a Goldstone boson \cite{Berezhiani:2015afa}, the neutron dark decay rate would be strongly suppressed and then for our benchmark choices of $m_{\chi}=938 \text{ MeV}$ and $m_{\phi}=0$ (see below) we would require $g_{\phi}$ to be larger by a factor of $2m_n/\left(m_n-m_{\chi}\right)\approx 1900$ in order to get a 1\% branching ratio} is a few times $10^{-14}$.  However, as the $\chi$ mass approaches the neutron mass, the phase space for the neutron to decay into vanishes, and thus the required coupling $g_{\phi}$ rises rapidly.  

There are a variety of constraints on the masses of the $\chi$ and $\phi$.  The neutron can only decay in this channel if $m_n>m_{\chi}+m_{\phi}$.  We also require that the proton is stable, so that the decay $p\rightarrow n+e^++\nu_e\rightarrow \chi+\phi+e^++\nu_e$ does not occur, and thus we demand that the sum of the masses of the final state exceeds the proton mass, that is, $m_{\chi}+m_{\phi}>m_p-m_e$.  However, this constraint is superseded by the requirement of the stability of $^9\text{Be}$, which requires $m_{\chi}+m_{\phi}>937.900\text{ MeV}$ \cite{Fornal:2023wji}.  Furthermore, thought not necessary, in this paper we will assume $m_{\chi}>m_{\phi}$.  

It would be elegant if the dark baryon that solves the neutron decay anomaly were to also make up all (or most) of the dark matter.  In this work, for convenience we will make assumptions that are inconsistent with what is known about dark matter \cite{Tulin:2017ara}, but we will keep in mind the possibility of the dark baryon being dark matter for future reference.  In such a case, we would need it to be stable, which leads to the requirement $m_{\chi}-m_{\phi}<m_p+m_e$.  In summary, we require
\begin{subequations}
\begin{align}
    937.900\text{ MeV} < &m_{\chi}+m_{\phi} < m_n\\
    &m_{\chi}-m_{\phi}<m_p+m_e.
\end{align}
\end{subequations}

Here, we will focus on the case\footnote{Actually, for this precise choice of parameters, it has been pointed out \cite{Berezhiani:2018eds,McKeen:2020zni} that the hydrogen atom becomes destabilized and can decay to $\chi+\phi+\nu$ on long timescales.  We could prevent this by raising the mass of the $\chi$ particle, but our results will not change meaningfully because in-medium effects operating inside neutron stars dominate over a small increase in the $\chi$ mass.} $m_{\chi}=938 \text{ MeV}$ and $m_{\phi}=0$.  In this case, there is relatively little phase space for the neutron dark decay channel, and thus a larger coupling $g_{\phi}$ is needed to achieve a 1\% branching ratio.  In this case, $g_{\phi}\approx 1.7\times 10^{-13}$, which is indicated in Fig.~\ref{fig:gphiplot}.  However, there is certainly room to examine even larger values of $g_{\phi}$ if one takes the $\chi$ mass to be increasingly close to the neutron mass.  Ultimately, $g_{\phi}$ is limited from above by the Raffelt criterion, discussed in Sec.~\ref{sec:rapid_n_decays}. 

A range of $g_{\phi}$ orders of magnitude smaller than those depicted in Fig.~\ref{fig:gphiplot} is also excluded by the existence of very cold neutron stars \cite{McKeen:2020oyr}.  If $g_{\phi}$ were this small, the production of $\chi$ particles would be so slow that even in an old neutron star, equilibrium would not yet have been reached, and thus the heat generated by ongoing neutron decays would heat the star beyond current constraints.  However, the choices of $g_{\phi}$ considered in this work correspond to $\chi$ production timescales much shorter than even a day (see Sec.~\ref{sec:results}).  Similar constraints are placed on neutron-mirror neutron oscillations \cite{Goldman:2019dbq,Berezhiani:2020zck,McKeen:2021jbh}.

If we had instead used the 0.15\% branching ratio bound from Dubbers \textit{et al.}~\cite{Dubbers:2018kgh}, then we would have obtained a value of $g_{\phi}$ that is smaller by about a factor of two and a half (which, if desired, could be counteracted by increasing the value of $m_{\chi}$ so it is even closer to the neutron mass.)  As will be seen in the results in Sec.~\ref{sec:results}, such a decrease in $g_{\phi}$ will have essentially no effect on our results, as this change is dwarfed by other, in-medium, effects inside neutron stars.
%%%%%%%%%%%%%%%%%%%%%%%%%%%%%%%%%%%%%%%%%%%%%%%%%%%
\section{$npe\chi$ matter}\label{sec:npechi_matter}
The IUF EoS \cite{Fattoyev:2010mx}, a relativistic mean field theory (RMF) \cite{Glendenning:1997wn}, is a highly successful model of dense matter, matching closely with terrestrial experiments near nuclear saturation density \cite{Dutra:2014qga}.  Recently, the EoS was modified slightly by Nandi \textit{et al.}~\cite{Nandi:2018ami} to give it additional repulsion at high density, raising its maximum neutron star mass from $1.96M_{\odot}$ to $2.08M_{\odot}$, which is consistent with the observed high-mass neutron stars.  We use this extension, called IUF-II \cite{Nandi:2018ami}, as our model of $npe$ matter.

\begin{table*}[]
\begin{tabular}{ccccccccccc}
\hline
Model & $m_{\sigma}$ {[}MeV{]} & $m_{\omega}$ {[}MeV{]} & $m_{\rho}$ {[}MeV{]} & $g_{\sigma}^2$ & $g_{\omega}^2$ & $g_{\rho}^2$ & $\tilde{\kappa}$ & $\tilde{\lambda}$ & $\tilde{\zeta}$ & $\Lambda_v$ \\ \hline
IUF-II & 491.5 & 782.5 & 763.0 & 97.1460 & 163.3050 & 184.6877 & 3.961721 & -0.006997 & 0.020 & 0.046 \\ \hline
\end{tabular}
\caption{Parameters in the $npe$ matter IUF-II EoS (c.f.~Eq.~\ref{eq:L_npe_int}) \cite{Nandi:2018ami}.}
\label{tab:IUFII_parameters}
\end{table*}

The IUF-II EoS has the interaction Lagrangian
\begin{align}
\mathcal{L}_{npe}^{\text{int}} &= g_{\sigma}\bar{\psi}\psi\sigma -g_{\omega}\bar{\psi}\gamma^{\mu}\psi\omega_{\mu}-\dfrac{1}{2}g_{\rho}\bar{\psi}\gamma^{\mu}\overrightarrow{\tau}\cdot\overrightarrow{\rho}_{\mu}\psi \nonumber\\
&-\dfrac{\tilde{\kappa}}{3!}g_{\sigma}^3\sigma^3-\dfrac{\tilde{\lambda}}{4!}g_{\sigma}^4\sigma^4+\dfrac{\tilde{\zeta}}{4!}g_{\omega}^4\left(\omega_{\mu}\omega^{\mu}\right)^2\nonumber\\
&+\Lambda_{\omega}g_{\omega}^2\overrightarrow{\rho}_{\mu}\cdot\overrightarrow{\rho}^{\mu}\omega_{\nu}\omega^{\nu}.\label{eq:L_npe_int}
\end{align}  
The parameters in the Lagrangian are given in Table \ref{tab:IUFII_parameters}.  The free Lagrangian $\mathcal{L}_{npe}^{\text{free}}$ contains the usual kinetic and mass terms for the neutron, proton, and electron (which is itself treated as a free Fermi gas).  
%%%%%%%%%%%%%%%%%%%%%%%%%%%%%%%%%%%%%%%%%%%%%%%%%%%%%%%
\subsection{Self-interacting dark baryons}
Given the existence of two new particles, $\chi$ and $\phi$, we consider their role in equilibrium properties of neutron stars.  As mentioned in the introduction, we now face a fork in the road.  Do the dark matter particles form a fluid of their own, and if so, is the neutron star as a whole two fluids or one?  That is, does the dark matter fluid interact strongly enough with the $npe$ matter to couple the fluids inextricably, or are the two fluids weakly coupled to one another?  The answer clearly depends on the rate of $n\rightarrow\chi+\phi$ in medium, but other factors could come into the picture as well.

First, we assume the $\phi$ particles, which we consider to be massless, have a long mean free path and free-stream from the system, just as neutrinos do from cold neutron stars.  As they are massless, gravitational trapping of the $\phi$ particles does not occur \cite{Caputo:2022mah}.   Absorption of the $\phi$ particles cannot be ignored in the case where the coupling $g_{\phi}$ is very large, but we will see later that if the neutron decay anomaly is solved, then $g_{\phi}$ is small enough that we can neglect absorption.  Next, we will assume the $\chi$ particles self-interact, though we motivate this choice later.  That self-interaction can be quite strong even if the $\chi$ is assumed to be dark matter \cite{Tulin:2017ara}.  Finally, how strongly are the two fluids coupled?  If the $\chi$ is dark matter, the bounds on the nucleon-dark matter cross section demand a weak coupling between the two fluids.  However, a two-fluid system makes bulk viscosity calculations complicated \cite{Gusakov:2007px}, so for simplicity we will assume that the $npe$ matter and $\chi$ fluid are locked together as one fluid, and hereafter discuss the EoS of $npe\chi$ matter, deferring the two-fluid scenario to future work.  However, we assume the thermal equilibration mechanism that locks the two components together is not the neutron dark decay $n\rightarrow\chi+\phi$, but instead some unspecified elastic scattering process $n+\chi\rightarrow n+\chi$.

The EoS of the dark baryons is that of a Fermi gas with self-repulsion, which is necessary to hold up a $2M_{\odot}$ neutron star.  Borrowing from the RMF framework, we generate the self-repulsion between the $\chi$ particles by having them exchange a dark vector meson, the $\omega'$.  Clearly, the interaction Lagrangian is 
\begin{equation}
    \mathcal{L}_{\chi}^{\text{int.}} = -g_{\omega'}\bar{\chi}\gamma^{\mu}\chi\omega'_{\mu}.  
\end{equation}
In infinite matter, only the ratio of the $\omega'$ mass to the coupling $g_{\omega'}$ matters \cite{Glendenning:1997wn}; we denote this quantity
\begin{equation}
    G' \equiv \left(\dfrac{g_{\omega'}}{m_{\omega\prime}}\right)^2,\label{eq:Gprime}
\end{equation}
and we keep it as a free parameter in this work.  The relevant equation of motion, in the mean field approximation, for the dark baryons is
\begin{equation}
    g_{\omega'}\omega'_0=G'n_{\chi}.
\end{equation}
We could have chosen to also allow the dark baryons to interact via a scalar meson exchange in addition to the vector exchange.  The scalar exchange is attractive, and would counteract the stiffening caused by the dark vector exchange, requiring us to further increase the strength of the dark vector repulsion in order to enable the EoS to support a $2M_{\odot}$ neutron star.  For simplicity, we solely consider the $\chi$ repulsive self-interaction.

The pressure and energy density of the dark baryons are
\begin{subequations}
\begin{align}
    P_{\chi} &= P_{\text{kinetic}}^{\chi} + \dfrac{1}{2}G'n_{\chi}^2\\
    \varepsilon_{\chi} &= \varepsilon^{\chi}_{\text{kinetic}} + \dfrac{1}{2}G'n_{\chi}^2,
\end{align}
\end{subequations}
where
\begin{subequations}
\begin{align}
    P_{\text{kinetic}}^{\chi} &\equiv \dfrac{2}{3}\int_0^{\infty}\dfrac{\mathop{d^3k}}{\left(2\pi\right)^3}\dfrac{k^2}{\sqrt{k^2+m_{\chi}^2}}f_{\chi}\\
    \varepsilon^{\chi}_{\text{kinetic}} &= 2\int_{0}^{\infty}\dfrac{\mathop{d^3k}}{\left(2\pi\right)^3}\sqrt{k^2+m_{\chi}^2}f_{\chi},
\end{align}
\end{subequations}
where $f$ is the Fermi-Dirac distribution function for the $\chi$ particles.

Because the dark baryons interact feebly (compared to the strong interaction strength) with standard model particles, even if the interaction is strong enough to allow for the consideration of one $npe\chi$ fluid, the dark baryons contribute additively to the energy density and pressure of the $npe\chi$ matter - any interaction term is negligible.  Interaction terms were considered in other works, however \cite{Bastero-Gil:2024kjo,Divaris:2025ftf}.   
%%%%%%%%%%%%%%%%%%%%%%%%%%%%%%%%%%%%%%%%%%%%
\subsection{Constraints on dark baryon self-repulsion}
\begin{figure}\centering
\includegraphics[width=0.45\textwidth]{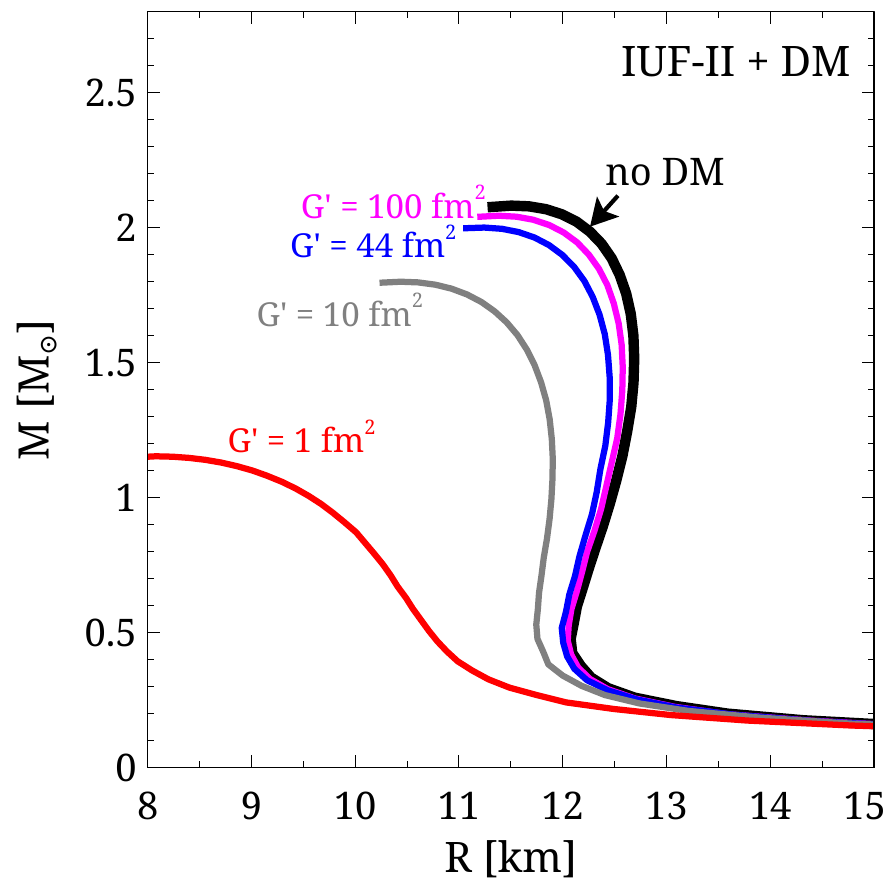}
\caption{Mass-radius curve of $npe\chi$-matter neutron stars with various values of $\chi$ self-repulsion strength $G'$.  The pure $npe$ case, with no dark baryons, is shown in black.  The $npe$ EoS can be achieved by taking the limit $G'\rightarrow\infty$.}
\label{fig:MR_curves}
\end{figure}

\begin{table}[]
\begin{tabular}{ccccl}
\cline{1-4}
\multicolumn{1}{|c|}{Model}  & \multicolumn{1}{c|}{$G_{\sigma}$} & \multicolumn{1}{c|}{$G_{\omega}$} & \multicolumn{1}{c|}{$G_{\rho}$} &  \\ \cline{1-4}
\multicolumn{1}{|c|}{IUF-II} & \multicolumn{1}{c|}{15.98}        & \multicolumn{1}{c|}{10.76}        & \multicolumn{1}{c|}{12.31}      &  \\ \cline{1-4}
\multicolumn{1}{|c|}{NL3}    & \multicolumn{1}{c|}{15.68}        & \multicolumn{1}{c|}{10.49}        & \multicolumn{1}{c|}{5.303}      &  \\ \cline{1-4}
\multicolumn{1}{l}{}         & \multicolumn{1}{l}{}              & \multicolumn{1}{l}{}              & \multicolumn{1}{l}{}            & 
\end{tabular}
\caption{Values of the quantity $G_i \equiv \left(g_i/m_i\right)^2$ (in $\text{fm}^2$) for each of the two EoSs discussed in this work.}
\label{table:Gvalues}
\end{table}

The strength of the repulsion between $\chi$ particles is not known and we treat it as a free parameter in this work.  However, it is not unconstrained.  The only hard bound we consider on $G'$ comes from the requirement that our $npe\chi$ EoS allows for stable neutron stars with masses of at least $2.0M_{\odot}$, as these have been observed (see the summary in \cite{Koehn:2024set}).  For small values of $G'$, there is very little energy cost to producing $\chi$ particles, and the neutron star will contain many of them, unacceptably softening the EoS.  As the repulsion strength is increased, the penalty for dark baryons increases, fewer of them are present, and the EoS stiffens.  Eventually, for sufficiently large $G'$, the neutron star will have very few $\chi$s and thus the EoS will be essentially the $npe$ EoS (in our case, IUF-II).  In this model, dark baryons can only soften the EoS, and the stiffest EoS is that without any dark baryons present.  The effect of the $\chi$ repulsion on the mass-radius curves is illustrated in Fig.~\ref{fig:MR_curves}.  As long as $G'\geq 44\text{ fm}^2$, the $2M_{\odot}$ requirement is met.  For comparison, in Table \ref{table:Gvalues} we show the analogous scalar attraction and vector repulsion strengths for the $npe$ matter EoSs used in this work.

\begin{figure*}\centering
\includegraphics[width=0.45\textwidth]{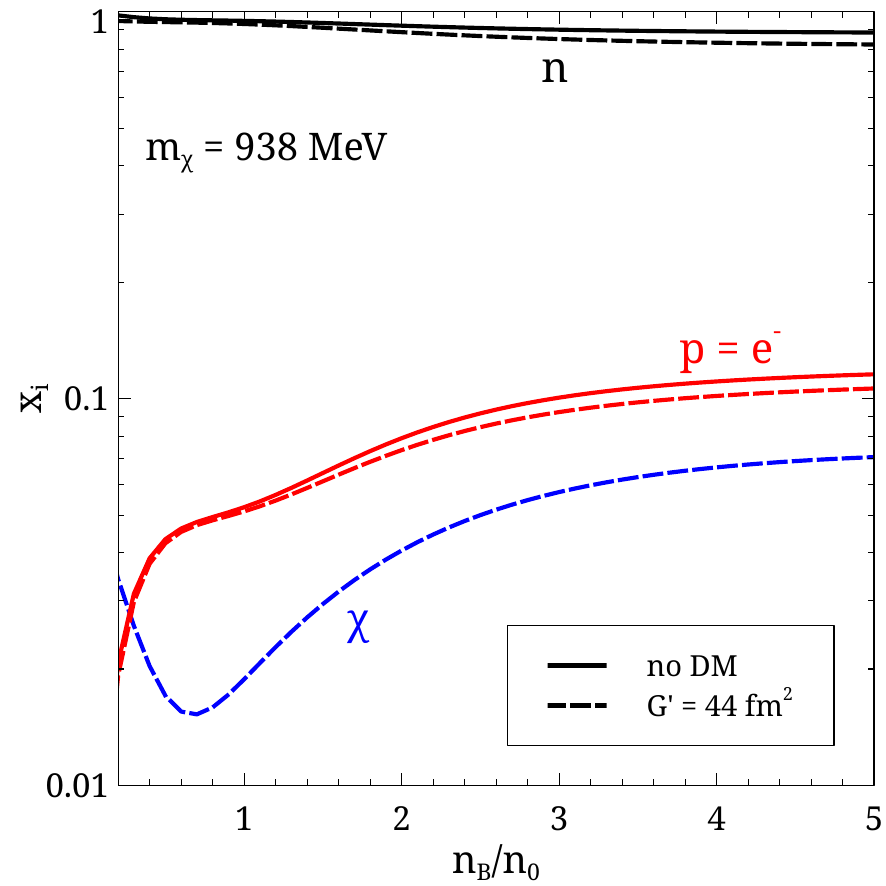}\hspace{1cm}
\includegraphics[width=0.45\textwidth]{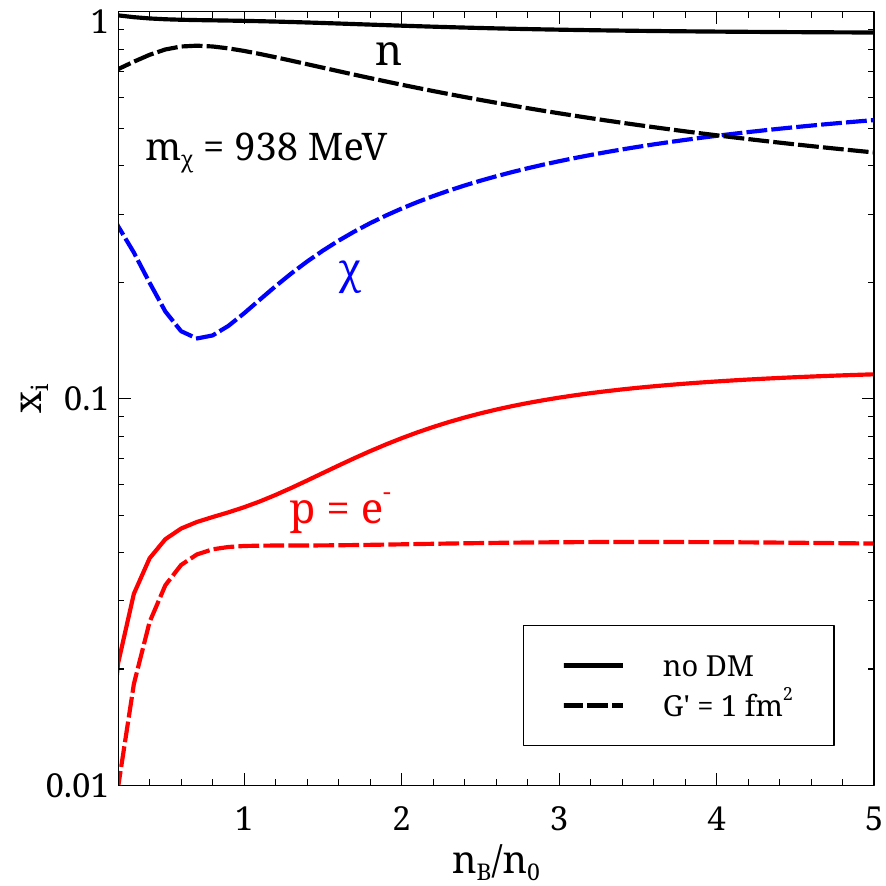}
\caption{Particle fractions in beta equilibrium for self-repulsion strengths of $G'=44\text{ fm}^2$ (the minimum value allowed by the existence of $2M_{\odot}$ stars, left panel) and $G' = 1\text{ fm}^2$ (right panel).  The particle fractions are calculated at zero temperature.  The baryon density $n_B$ includes the dark baryon content (c.f.~Eq.~\ref{eq:nB_definition}).}
\label{fig:particlefractions}
\end{figure*}

Fig.~\ref{fig:particlefractions} illustrates the beta-equilibrium values of the particle fractions as a function of the baryon density $n_B$, which includes the dark baryons
\begin{equation}
    n_B\equiv n_n+n_p+n_{\chi}.\label{eq:nB_definition}
\end{equation}
Adding dark baryons to the EoS lowers the neutron content of the system, as well as the proton and electron fractions.  For the lowest value of the repulsion strength that supports a $2M_{\odot}$ neutron star, $G'=44\text{ fm}^2$, the dark baryons make up a few percent of the baryon content (see the left hand panel of Fig.~\ref{fig:particlefractions}).  If the repulsion strength is larger, than even fewer dark baryons are present.  In the right hand panel of Fig.~\ref{fig:particlefractions}, we show the particle fractions in beta equilibrium for $G'=1\text{ fm}^2$, just to show how large the $\chi$ content becomes.  Of course, a stiffer $npe$ matter EoS would allow for a higher dark baryon content while still satisfying the $2M_{\odot}$ requirement.

If one wants to consider the dark baryons as making up all of the dark matter, then additional constraints on $G'$ come into play.  The Born approximation cross section, in the nonrelativistic limit, for $\chi-\chi$ scattering via the exchange of a (heavy) dark vector meson is \cite{Girmohanta:2022dog}
\begin{equation}
    \sigma_{\chi\chi} \approx \dfrac{G'^2m_{\chi}^2}{8\pi}.\label{eq:sigma_chichi}
\end{equation}
Often in cosmology, the quantity that can be constrained is $\sigma_{\chi\chi}/m_{\chi}$, which can be written
\begin{equation}
    \dfrac{\sigma_{\chi\chi}}{m_{\chi}}=5.75\times 10^{-3} \,\dfrac{\text{cm}^2}{\text{g}}\times\left(\dfrac{G'}{\text{fm}^2}\right)^2\left(\dfrac{m_{\chi}}{\text{GeV}}\right).
\end{equation}
Originally, dark matter was treated as collisionless, but recently, motivation has emerged for self-interacting dark matter.  This development is reviewed in \cite{Tulin:2017ara,Adhikari:2022sbh}.   For example, collisionless dark matter simulations of dark matter halos produce a dark matter density profile that is ``cuspy'', that is, one that continues to rise as one moves toward the center of the halo.  However, rotation curves indicate that the dark matter profile should level off at the center (forming a dark matter ``core'').  Giving the dark matter a self-interaction cross section of $\sigma_{\chi\chi}/m_{\chi}\sim 1 \text{ cm}^2/\text{g}$ allows for significantly increased thermalization of the dark matter core and yields a dark matter density profile in line with observations. Many other simulations also find that dark matter cross sections within an order of magnitude of  $1 \text{ cm}^2/\text{g}$ solve other inconsistencies that exist in the collisionless dark matter paradigm.  However, there is still considerable uncertainty as cosmological simulations often do not include baryonic (non dark) matter at all.  In addition, the scattering cross section could have velocity or angular dependence that is not typically accounted for.  

In any case, while we do not demand that the $\chi$ is dark matter (it could be just some extra degree of freedom that exists within the high density environment of neutron stars), if one wants to think of it as such, in order to support a $2M_{\odot}$ neutron star, $\sigma_{\chi\chi}/m_{\chi}$ must be at least $10 \text{ cm}^2/\text{g}$, which is on the high side for self-interacting dark matter.  However, this minimum value depends on the underlying nuclear EoS considered (see \cite{Das:2025pjl,Shirke:2023ktu}).
%%%%%%%%%%%%%%%%%%%%%%%%%%%%%%%%%%%%%%%%%%%%%
\subsection{Thermodynamics of $npe\chi$ matter}
Here, we derive thermodynamic relationships in $npe\chi$ matter, where we do not impose beta equilibrium, instead allowing the system to be out of beta equilibrium in two different (and independent) ``directions"
\begin{subequations}
\begin{align}
    \delta\mu_1&\equiv \mu_n-\mu_p-\mu_e\label{eq:beq1}\\
    \delta\mu_2 &\equiv \mu_n-\mu_{\chi}.\label{eq:beq2}
\end{align}
\end{subequations}
Beta equilibrium occurs when $\delta\mu_1=0$ and $\delta\mu_2=0$.  Actually, this is only true at zero temperature, as will be discussed at the end of this section.  

Just as there are two independent beta equilibrium conditions, there are two independent particle fractions which we choose to be $x_p$ and $x_{\chi}$ ($x_i\equiv n_i/n_B$).  The Euler equation can be written
\begin{equation}
    \dfrac{\varepsilon+P}{n_B}=\dfrac{s}{n_B}T+\mu_n-x_p\delta\mu_1-x_{\chi}\delta\mu_2
\end{equation}
and the first law of thermodynamics is
\begin{equation}
    \mathop{d\left(\frac{\varepsilon-sT}{n_B}\right)}=\frac{P}{n_B^2}\mathop{dn_B}-\frac{s}{n_B}\mathop{dT}-\delta\mu_1\mathop{dx_p}-\delta\mu_2\mathop{dx_{\chi}}.
\end{equation}
From this equation, we can derive a few relevant Maxwell relations,
\begin{subequations}
\begin{align}
    \dfrac{\partial P}{\partial x_p}\bigg\vert_{n_B,x_{\chi},T}&=-n_B^2\dfrac{\partial \delta\mu_1}{\partial n_B}\bigg\vert_{x_p,x_{\chi},T}\\
    \dfrac{\partial P}{\partial x_{\chi}}\bigg\vert_{n_B,x_{p},T}&=-n_B^2\dfrac{\partial \delta\mu_2}{\partial n_B}\bigg\vert_{x_p,x_{\chi},T}\\
     \dfrac{\partial \delta\mu_1}{\partial x_{\chi}}\bigg\vert_{n_B,x_p,T}&=\dfrac{\partial \delta\mu_2}{\partial x_p}\bigg\vert_{n_B,x_{\chi},T}.
\end{align}
\end{subequations}
Finally, for future use, we define the susceptibilities
\begin{subequations}
\begin{align}
    A_i &\equiv n_B \dfrac{\partial \delta\mu_i}{\partial n_B}\bigg\vert_{x_p,x_{\chi},T},\label{eq:suscA}\\
    B_i &\equiv \dfrac{1}{n_B}\dfrac{\partial\delta\mu_i}{\partial x_p}\bigg\vert_{n_B,x_{\chi},T},\label{eq:suscB}\\
    C_i &\equiv \dfrac{1}{n_B}\dfrac{\partial\delta\mu_i}{\partial x_{\chi}}\bigg\vert_{n_B,x_p,T},\label{eq:suscC}
\end{align}
\end{subequations}
where $i=1,2$.  The Maxwell relations then become 
\begin{subequations}
\begin{align}
A_1 &= -\dfrac{1}{n_B}\dfrac{\partial P}{\partial x_p}\bigg\vert_{n_B,x_{\chi},T}\\
A_2 &= -\dfrac{1}{n_B}\dfrac{\partial P}{\partial x_{\chi}}\bigg\vert_{n_B,x_p,T}\\
C_1 &= B_2.\label{eq:maxwell_relation}
\end{align}
\end{subequations}
Finally, we introduce the isothermal compressibility of $npe\chi$ matter \cite{CompOSECoreTeam:2022ddl}
\begin{equation}\label{eq:compressibility}
    \kappa_T = \left( \left. n_B\dfrac{\partial P}{\partial n_B}\right\vert_{T,x_p,x_{\chi}}\right)^{-1}.
\end{equation}

In this work, we consider matter that is both transparent to neutrinos and to $\phi$ particles.  In such a case, $\delta\mu_1=0$ and $\delta\mu_2=0$ (as defined in Eqs.~\ref{eq:beq1} and \ref{eq:beq2}) are no longer the beta equilibrium conditions at finite temperature, as can be verified through explicit computation of the rate of proton or $\chi$ production \cite{Alford:2018lhf,Alford:2021ogv}.  Instead, the correct ``finite-temperature'' beta equilibrium conditions are
\begin{subequations}
\begin{align}
\mu_n &= \mu_p + \mu_e + \mu_{\delta,1}\label{eq:truebeq_1}\\
\mu_n &= \mu_{\chi} + \mu_{\delta,2},\label{eq:truebeq_2}
\end{align}
\end{subequations}
where $\delta\mu_1$ and $\delta\mu_2$, which depend on density and temperature, are found by adjusting their value away from zero until the net rates of $x_p$ and $x_{\chi}$ are each zero (as this is the only sensible definition of beta equilibrium).  As the temperature goes to zero, the correction terms $\mu_{\delta,1}$ and $\mu_{\delta,2}$ go to zero and the ``zero-temperature'' beta equilibrium conditions (Eqs.~\ref{eq:beq1} and \ref{eq:beq2}, set equal to zero) hold.
%%%%%%%%%%%%%%%%%%%%%%%%%%%%%%%%%%%%%%%%%%%%%%%%%%%%%
\section{Bulk viscosity in $npe\chi$ matter}\label{sec:bulkviscosity}
In this section, we derive the bulk viscosity by considering a fluid element of $npe\chi$ matter undergoing a periodic density oscillation, causing it to be pushed out of chemical equilibrium.  We will sketch the derivation, which follows almost exactly the derivations of bulk viscosity in $npe$ matter \cite{Harris:2020rus,Harris:2024evy}, $npe\mu$ matter \cite{Harris:2024evy}, and in neutrino-trapped $npe\mu\pi$ matter \cite{Harris:2024ssp}.  We consider a fluid element undergoing a small amplitude, sinusoidal density oscillation
\begin{align}
    n_B(t) &= n_B + \delta n_B \cos{(\omega t)},
\end{align}
where $\delta n_B \ll n_B$ and we have chosen $\delta n_B$ to be real. The density oscillation pushes the particle fractions out of beta equilibrium
\begin{align}
    x_i(t) &= x_i^0 + \Re{(\delta x_i)}\cos{(\omega t)} - \Im{(\delta x_i)}\sin{(\omega t)}\label{eq:xi_of_t}
\end{align}
and since the pressure depends on the baryon density and the particle fractions, it oscillates harmonically too
\begin{align}\label{eq:P_of_t}
    P(t) &= P_0 +\Re{(\delta P)}\cos{(\omega t)} - \Im{(\delta P)}\sin{(\omega t)}.
\end{align}
The imaginary parts of the pressure and particle fractions represent the degree to which they oscillate out of phase with the baryon density, which is the effect that gives rise to bulk viscosity.  The bulk viscosity coefficient is given by \cite{Harris:2020rus,Harris:2024evy}
\begin{equation}\label{eq:zeta_imP}
    \zeta = \left(\dfrac{n_B}{\delta n_B}\right)\dfrac{\Im{(\delta P)}}{\omega}.
\end{equation}

The pressure is a function $P=P\left(n_B,T,x_p,x_{\chi}\right)$.  For convenience, we will consider isothermal (as opposed to adiabatic) oscillations.  Therefore, the pressure during an oscillation can be expressed as
\begin{align}
    P &= P_0 + \dfrac{\partial P}{\partial n_B}\bigg\vert_{T,x_p,x_{\chi}}\Re{(\delta n_B e^{i\omega t})}\label{eq:P_taylor_Exp}\\
    &+\dfrac{\partial P}{\partial x_p}\bigg\vert_{T,n_B,x_{\chi}}\Re{(\delta x_p e^{i\omega t})}+\dfrac{\partial P}{\partial x_{\chi}}\bigg\vert_{T,n_B,x_p}\Re{(\delta x_{\chi} e^{i\omega t})}\nonumber\\
    &= P_0 + \dfrac{\kappa_T^{-1}}{n_B}\Re{(\delta n_B e^{i\omega t})}-n_BA_1\Re{(\delta x_p e^{i\omega t})}\nonumber\\
    &-n_BA_2\Re{(\delta x_{\chi} e^{i\omega t})}.\nonumber
\end{align}
Therefore,
\begin{align}
    \Im{(\delta P)} &= -n_BA_1\Im{(\delta x_p)}-n_BA_2\Im{(\delta x_{\chi})}
\end{align}
and thus the bulk viscosity is
\begin{equation}
    \zeta = -\dfrac{n_B^2}{\delta n_B}\dfrac{1}{\omega}\left[A_1\Im{(\delta x_p)}+A_2\Im{(\delta x_{\chi})}\right].\label{eq:bv_nextstep}
\end{equation}
Now we find expressions for the $\Im{(\delta x_{i})}$.

The particle fractions, at fixed volume, evolve due to chemical reactions according to
\begin{subequations}
\begin{align}
    n_B\dfrac{\mathop{dx_p}}{\mathop{dt}} &= \Gamma_{n\rightarrow p}-\Gamma_{p\rightarrow n} \approx \lambda_1\delta\mu_1,\label{eq:nbdxp}\\
    n_B\dfrac{\mathop{dx_{\chi}}}{\mathop{dt}} &=\Gamma_{n\rightarrow\chi}-\Gamma_{\chi\rightarrow n} \approx \lambda_2\delta\mu_2.\label{eq:nbdxchi}
\end{align}
\end{subequations}

Neutrons can be turned into protons via the standard neutron decay processes
\begin{subequations}
\begin{align}
    n &\rightarrow p + e^- + \bar{\nu}_e \quad \text{direct Urca}\\
    n + n &\rightarrow p + e^- + \bar{\nu}_e + n \quad \text{modified Urca}
\end{align}
\end{subequations}
and protons can be turned into neutrons via electron capture
\begin{subequations}
\begin{align}
    e^- + p &\rightarrow n + \nu_e \quad \text{direct Urca}\\
    e^- + p + n &\rightarrow n + \nu_e + n \quad \text{modified Urca}.
\end{align}
\end{subequations}
We neglect the possibility of $\chi$ as a spectator particle, because direct Urca is the dominant Urca process in the high-temperature environments discussed in this paper \cite{Alford:2018lhf}.  The rate $\Gamma_{n\rightarrow p}$ is the direct Urca neutron decay rate plus the modified Urca neutron decay rate (plus, in principle, the neutron decay rates accounting for multiple spectator particles).  The opposing rate $\Gamma_{p\rightarrow n}$ for electron capture has the analogous decomposition.

Neutrons can be turned into dark baryons via the processes 
\begin{subequations}
\begin{align}
    n &\rightarrow \chi + \phi \quad \text{direct}\\
    n + n &\rightarrow \chi + \phi + n \quad \text{modified (n-spec)}\\
    n + \chi &\rightarrow \chi + \phi + \chi \quad \text{modified } (\chi\text{-spec)}
\end{align}
\end{subequations}
and dark baryons can be turned into neutrons via
\begin{subequations}
\begin{align}
    \chi &\rightarrow n + \phi \quad \text{direct}\\
    \chi + n &\rightarrow n + \phi + n \quad \text{modified (n-spec)}\\
    \chi + \chi &\rightarrow n + \phi + \chi \quad \text{modified } (\chi\text{-spec)}
\end{align}
\end{subequations}
We include the dark baryon as a possible spectator in the neutron dark decay channel because the direct process is suppressed, as will be shown later.  As with Urca, the rate $\Gamma_{n\rightarrow \chi}$ is the direct neutron dark decay rate plus the modified neutron dark decay rates ($n$ and $\chi$ spectators) plus higher order diagrams, however we will use the nucleon width approximation instead of this decomposition (see Sec.~\ref{sec:neutron_dark_decay}).

As indicated in Eqs.~\ref{eq:nbdxp} and \ref{eq:nbdxchi}, while the net rate $\overrightarrow{\Gamma}-\overleftarrow{\Gamma}$ can be calculated for arbitrary deviations from equilibrium $\delta\mu$, in this paper we study only \textit{subthermal} bulk viscosity \cite{Haensel:2002qw} where the deviation from chemical equilibrium $\delta\mu_i\ll T$, and thus the beta equilibration rates are given by their linearized forms $\lambda$.  For the rest of this section, we assume arbitrary $\lambda$, deferring a calculation of the rates to a later section.  

As the baryon density oscillates around some background density, the particle fractions oscillate around their beta equilibrium values, with some phase lag, as expressed in Eq.~\ref{eq:xi_of_t}.  Another way of expressing this departure from beta equilibrium is via tracking the chemical potential differences $\delta\mu_i$, which can can be written
\begin{align}
    \delta\mu_i &= A_i\dfrac{\delta n_B}{n_B}\cos{(\omega t)} + n_B \big[B_i \Re{(\delta x_p e^{i\omega t})}+C_i\Re{(\delta x_{\chi} e^{i\omega t})}\big]\nonumber\\
    &=A_i\dfrac{\delta n_B}{n_B}\cos{(\omega t)}+ n_{B}\left[B_i\Re{(\delta x_p)}+C_i\Re{(\delta x_{\chi})}\right]\cos{\left(\omega t\right)}\nonumber\\
    &-n_{B}\left[B_i\Im{(\delta x_p)}+C_i\Im{(\delta x_{\chi})}\right]\sin{\left(\omega t\right)}.\label{dmu_expanded}
\end{align}
Using Eqs.~\ref{eq:xi_of_t} and \ref{dmu_expanded}, with Eq.~\ref{eq:bv_nextstep}, and using the Maxwell relation Eq.~\ref{eq:maxwell_relation} to replace $C_1$ with $B_2$, we find that the subthermal bulk viscosity of $npe\chi$ matter is given by
\begin{widetext}
\begin{equation}
    \zeta = \dfrac{\lambda_1\lambda_2\left[\left(A_2B_1-A_1B_2\right)^2\lambda_1+\left(A_2B_2-A_1C_2\right)^2\lambda_2\right]+\left(A_1^2\lambda_1+A_2^2\lambda_2\right)\omega^2}{\left(B_2^2-B_1C_2\right)^2\lambda_1^2\lambda_2^2+\left(B_1^2\lambda_1^2+2B_2^2\lambda_1\lambda_2+C_2^2\lambda_2^2\right)\omega^2+\omega^4}.\label{eq:full_bv_formula}
\end{equation}
\end{widetext}
As discussed in \cite{Harris:2024evy,Harris:2024ssp}, while the bulk viscosity of a system with two equilibrating quantities ($\delta\mu_1$ and $\delta\mu_2$) is not, in general, separable into two individual pieces, it can still be useful to consider the ``partial'' bulk viscosities, which are the bulk viscosity with only one equilibration rate $\lambda_i$ active, with all other reaction rates $\lambda_{j\neq i}$ set to zero.  In the $npe\chi$ system, the partial bulk viscosities are
\begin{subequations}
\begin{align}
    \zeta_1 &= \left\vert\dfrac{A_1^2}{B_1}\right\vert\dfrac{\gamma_1}{\gamma_1^2+\omega^2},\\
    \zeta_2 &= \left\vert\dfrac{A_2^2}{C_2}\right\vert\dfrac{\gamma_2}{\gamma_2^2+\omega^2}
\end{align}
\end{subequations}
where we have defined the two equilibration rates
\begin{subequations}
\begin{align}
    \gamma_1 &\equiv \vert B_1\vert \lambda_1\label{eq:gamma1}\\
    \gamma_2 &\equiv \vert C_2\vert\lambda_2.\label{eq:gamma2}
\end{align}
\end{subequations}
$\gamma_1$ and $\gamma_2$ are the rates at which the proton and $\chi$ fractions (respectively) relax to their beta equilibrium values.  It can be shown, using Maxwell relations and the definitions of $A_i$, $B_i$, and $C_i$ (see Appendix H in \cite{Harris:2024ssp} and also \cite{Lindblom:2001hd}) that
\begin{align}
    \left\vert\dfrac{A_1^2}{B_1}\right\vert &= n_B\left(  \dfrac{\partial P}{\partial n_B}\bigg\vert_{x_{p},x_{\chi},T}- \dfrac{\partial P}{\partial n_B}\bigg\vert_{\delta\mu_1,x_{\chi},T}  \right)\label{eq:A1sqB1}\\
    &= n_B\left\vert \dfrac{\partial P}{\partial x_p}\bigg\vert_{n_B,x_{\chi},T} \dfrac{\partial x_p}{\partial n_B}\bigg\vert_{\delta\mu_1,x_{\chi},T} \right\vert\nonumber\\
    \left\vert\dfrac{A_2^2}{C_2}\right\vert &= n_B\left( \dfrac{\partial P}{\partial n_B}\bigg\vert_{x_{p},x_{\chi},T} - \dfrac{\partial P}{\partial n_B}\bigg\vert_{\delta\mu_2,x_{p},T} \right).\label{eq:A2sqC2}\\
    &= n_B\left\vert \dfrac{\partial P}{\partial x_{\chi}}\bigg\vert_{n_B,x_{p},T} \dfrac{\partial x_{\chi}}{\partial n_B}\bigg\vert_{\delta\mu_2,x_{p},T}\right\vert.\nonumber
\end{align}
Thus, the maximum value of one of the partial bulk viscosities at a fixed density and frequency $\omega$, which occurs when $\gamma_i=\omega$ (as long as the susceptibilities are relatively independent of temperature), depends on two factors.  One is the degree to which the pressure depends on the relevant particle fraction (for the Urca process, the proton fraction).  Of course, if the pressure does not depend at all on the particle fraction, no pdV work is done because the pressure does not change irreversibly and thus there is no bulk viscosity.  The second is the amount that the particle fraction's beta-equilibrium value varies with density (with all other particle fractions being fixed).  If the EoS dictates that the particle fraction in beta equilibrium is ``flat'' with respect to density, then little bulk viscosity is expected.  However, note that in a multicomponent system, one cannot just look at a figure like Fig.~\ref{fig:particlefractions} to determine the ``flatness'' of the derivative, because this figure is produced at fixed $\delta\mu_1$ and $\delta\mu_2$, that is, where all particle species are in beta equilibrium, while the derivative $\partial x_i/\partial n_B$ is at, say, fixed $\delta\mu_1$, but also at fixed $x_{\chi}$ (not fixed $\delta\mu_2$).  But for a single component system like $npe$ matter, this distinction does not exist and one could use Fig.~\ref{fig:particlefractions} to get a sense of $\partial x_i/\partial n_B\vert_{\delta\mu}$.
%%%%%%%%%%%%%%%%%%%%%%%%%%%%%%%%%%%%%%%%%%%%%%%%%%%%
\section{Results}\label{sec:results}
%%%%%%%%%%%%%%%%%%%%%%%%%%%%%%%%%%%%%%%%%%%%%%
\subsection{Urca bulk viscosity in $npe\chi$ matter with frozen dark baryons}
We will see in a later section that if $g_{\phi}$ is chosen so that the neutron dark decay has a 1\% branching ratio, then the beta equilibration rate $\gamma_2$ is much slower than the other scales in the problem, and can safely be taken to zero, and therefore the dark baryon content $x_{\chi}$ is (essentially) frozen during a density perturbation.  In this situation, the bulk viscosity $\zeta$ becomes exactly the partial bulk viscosity $\zeta_1$, as can be seen by setting $\lambda_2=0$ in Eq.~\ref{eq:full_bv_formula}.  Therefore, if the $\chi$ content is frozen, then the bulk viscosity of $npe\chi$ matter is the same as the bulk viscosity of $npe$ matter, except to the extent that the $\chi$ changes the EoS and its susceptibilities.  Thus, we will begin this section by studying how the presence of the $\chi$ particles in the EoS modifies the Urca bulk viscosity.

\begin{figure}\centering
\includegraphics[width=0.45\textwidth]{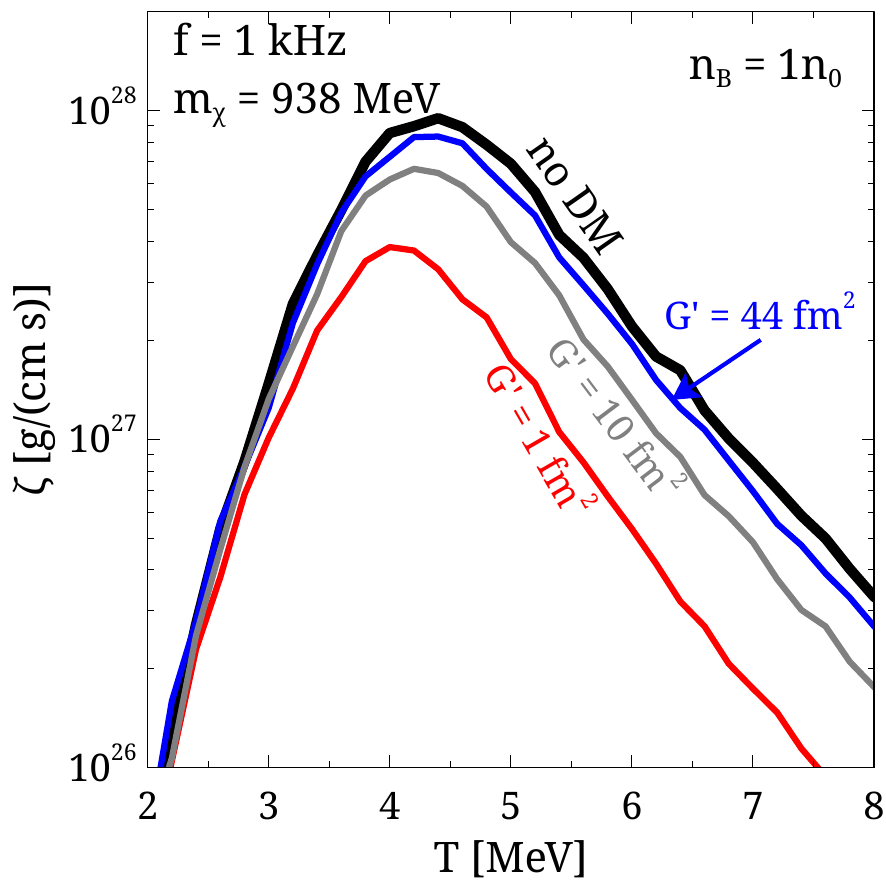}
\caption{Bulk viscosity of $npe\chi$ matter with frozen $\chi$ content (i.e., $\lambda_2=0$).  The density oscillation frequency is 1 kHz and the matter is at $n_B=1n_0$.  Different color curves correspond to different choices of the dark baryon self-repulsion strength.  The matter here is calculated in finite-temperature beta equilibrium (Eqs.~\ref{eq:truebeq_1} and \ref{eq:truebeq_2}).}
\label{fig:bulk_viscosity_l2zero}
\end{figure}

In Fig.~\ref{fig:bulk_viscosity_l2zero}, we plot the bulk viscosity $\zeta = \zeta_1$ with frozen dark baryons in $npe\chi$ matter subjected to a density oscillation with a linear frequency of 1 kHz.  The result without dark matter is plotted in black.  This result is consistent with previous calculations in the literature \cite{Harris:2024evy,Alford:2019qtm,Alford:2023gxq}, which show one resonant peak, with maximum value somewhere in the vicinity of $10^{28} \text{ g/(cm s)}$, with the peak occurring at a temperature around 3-5 MeV.  The peak location (temperature) at fixed density and oscillation frequency occurs where $\gamma(T)=\omega$ \cite{Harris:2024evy,Alford:2019qtm}.  The peak height is set by the susceptibilities of the EoS, which in the case of the Urca process in $npe$ matter, can be written in terms of the symmetry energy \cite{Alford:2010gw,Harris:2025ncu,Yang:2023ogo,Yang:2025yoo}.

To the extent that the dark baryons affect the Urca rate (through their effect on the EoS), the bulk-viscous peak moves to lower or higher temperatures.  To the extent that the dark baryons affect the susceptibilities $A_1$ and $B_1$, the peak shifts up and down.  Fig.~\ref{fig:bulk_viscosity_l2zero} indicates that when a large number of dark baryons are present (the $G'=1\text{ fm}^2$ case), the bulk viscosity viscosity can be significantly reduced.  However, for allowed values of the dark baryon self-repulsion strength ($G'\geq 44\text{ fm}^2$) the change is about a 30\% decrease, which is well within the uncertainty on the underlying $npe$ EoS bulk viscosity.  For an EoS like NL3 \cite{Lalazissis:1996rd}, which is much stiffer and allows for much greater dark baryon content (about 15\%, as opposed to a few percent for IUF-II) while still being consistent with $2M_{\odot}$ observations, the Urca bulk viscosity is reduced by a factor of 2-3 if the EoS contains dark baryons.  The dark baryons reduce the symmetry energy and its slope at saturation density (compared to the EoS without dark baryons), which reduces the bulk viscosity.  As the symmetry energy is sensitive to the number densities of neutrons and protons, being indifferent to dark baryons, at saturation density the neutron and proton densities sum to a little less than $n_0$ because some of the baryon number is stored in dark baryons.  Thus, the bulk viscosity maximum at $n_0$ probes the symmetry energy at a density less than $n_0$, which is why the dark baryons decrease the symmetry energy.  This discussion of the bulk viscosity was at $n_B=n_0$, but the story persists at higher densities too, as we will discuss now.  

\begin{figure*}\centering
\includegraphics[width=0.45\textwidth]{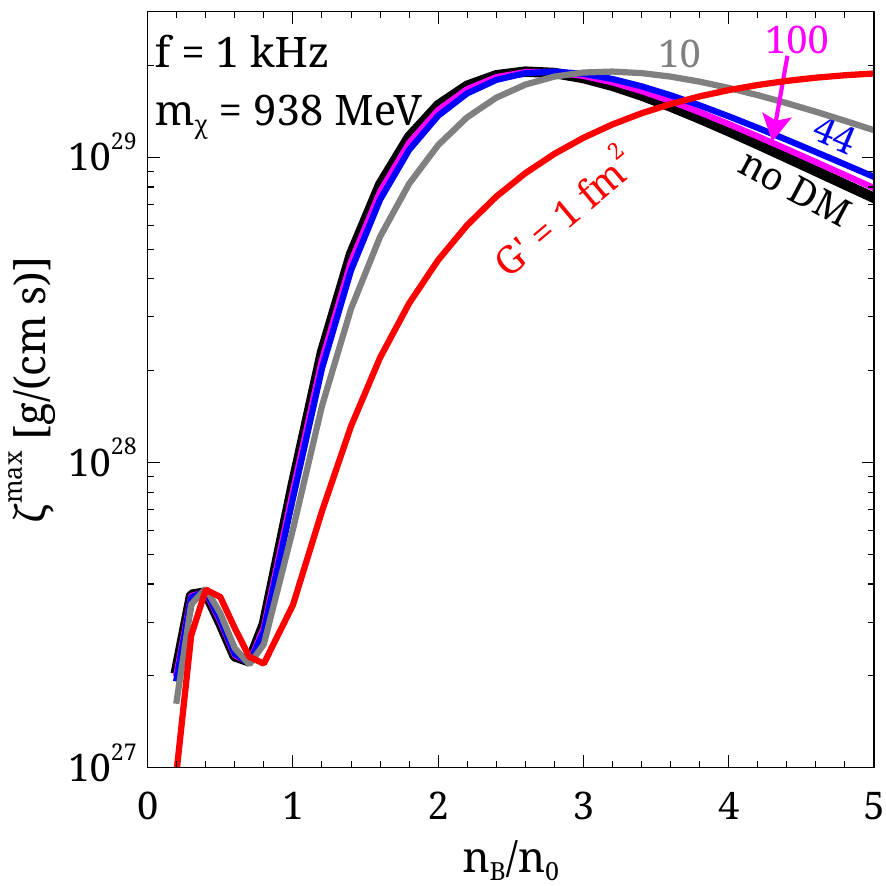}\hspace{1cm}
\includegraphics[width=0.45\textwidth]{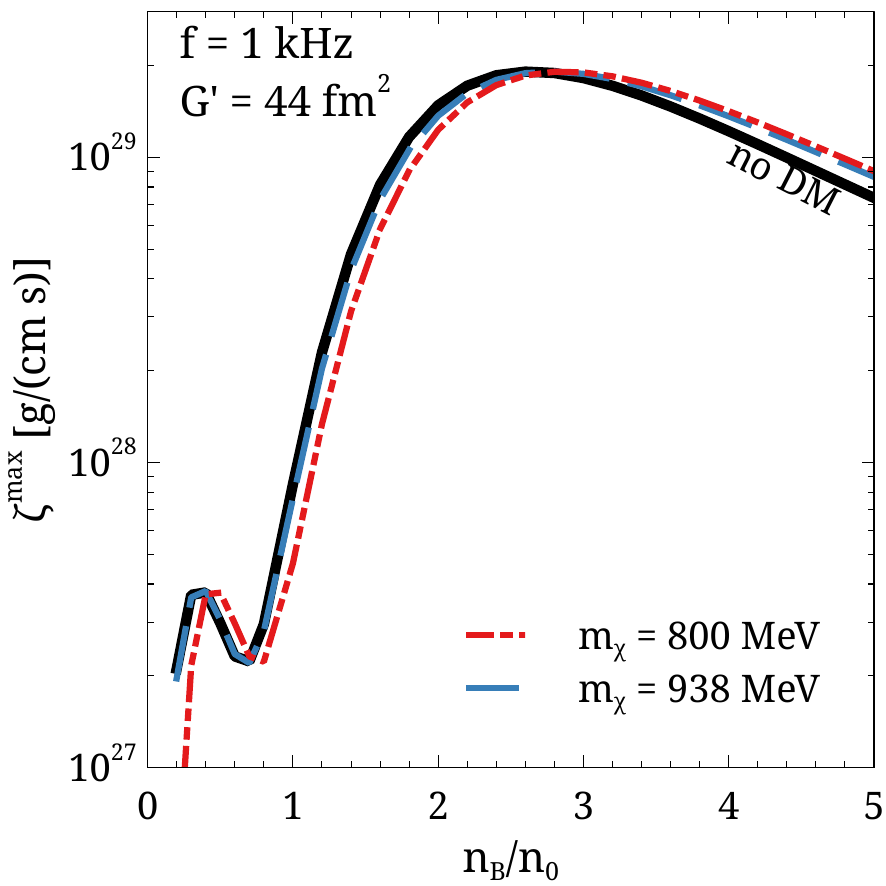}
\caption{Peak value (across all temperatures) of the Urca bulk viscosity $\zeta_1$, as a function of the baryon density.  The left panel shows a variety of dark baryon self-repulsion strengths $G'$, while the right panel shows different values of the dark baryon mass $m_{\chi}$.  The calculations in these plots are done at zero temperature.}
\label{fig:bvmax}
\end{figure*}

The bulk viscosity (still assuming $\lambda_2=0$) can be plotted as a contour plot in the density temperature plane (c.f.~Fig.~4 in \cite{Alford:2019qtm} or Fig.~4 in \cite{Alford:2023gxq}).  It will exhibit a ridge-like structure, peaking for the densities and temperatures where $\gamma_1(n_B,T)\approx\omega$.  One can think of, at a fixed density, the bulk viscosity having a peak at a particular temperature.  In Fig.~\ref{fig:bvmax}, we show the value of the resonant maximum of the bulk viscosity as a function of density: this essentially compresses a bulk viscosity contour plot into one dimension, following the ridge from low density to high density, allowing us to efficiently explore the dependence of the bulk viscosity on the parameters of the dark baryons.    

Without dark matter, Fig.~\ref{fig:bvmax} shows that IUF-II predicts the bulk viscous peak to rise with density, reaching its global maximum value at around $3n_0$.  At higher densities, it falls slowly.  This is consistent with the calculations done in \cite{Alford:2019qtm}, although with the IUF, not IUF-II EoS.  The behavior of the peak bulk viscosity with density is ultimately due to the density-dependence of the symmetry energy \cite{Harris:2025ncu,Yang:2023ogo,Yang:2025yoo}.  From the left panel of Fig.~\ref{fig:bvmax}, we see that for unacceptably low values of $\chi$ repulsion (like $G'=1\text{ fm}^2$), the curve $\zeta_{\text{max}}(n_B)$ has the same general shape as the case without dark baryons, but the maximum is shifted to much higher densities.  However, as the dark repulsion is increased sufficiently to hold up a $2M_{\odot}$ neutron star ($G'=44\text{ fm}^2)$, the bulk viscosity behavior becomes increasingly similar to the $npe$ behavior, deviating by, at most, 15\% (to be clear, at a specific density, the actual value of the bulk viscosity $\zeta(T)$ might deviate from the $npe$ bulk viscosity by more than this, but the difference in the peak value, at whatever temperature it occurs, with and without dark baryons is less than 15\%).  The right panel shows that the mass of the dark baryon has a small effect on the bulk viscosity for a given $G'$.  Similar trends hold for the NL3 EoS \cite{Lalazissis:1996rd}, where the dark baryons push the maximum to higher density, but at a given density, the maximum value of bulk viscosity changes by under a factor of two.

We note that Fig.~\ref{fig:bvmax} is calculated assuming the zero-temperature beta equilibrium conditions (Eqs.~\ref{eq:beq1} and \ref{eq:beq2}, set equal to zero).  By comparing with Fig.~\ref{fig:bulk_viscosity_l2zero} at $n_0$, we see that the two results are within 15\% of each other, so the peak value of the bulk viscosity does not change much when the beta equilibrium condition is corrected (Eq.~\ref{eq:truebeq_1} and \ref{eq:truebeq_2}).

Figs.~\ref{fig:bulk_viscosity_l2zero} and \ref{fig:bvmax} show that as long as the neutron dark decay is slow, the bulk viscosity of $npe\chi$ matter looks very much like the bulk viscosity of $npe$ matter: that is, the bulk viscosity has just one peak as a function of temperature, which occurs at $T\approx 3-5$ MeV. It is unlikely, given the current uncertainties in the $npe$ EoS due to our lack of understanding of the strong force, it is unlikely that the presence of dark baryons can be derived from bulk-viscous dissipation, as long as the neutron dark decays are very slow.  Thus, we now turn to calculating the neutron dark decay rate, but first we start with the Urca rate.  
%%%%%%%%%%%%%%%%%%%%%%%%%%%%%%%%%%%%%%%%%%%%%%
\subsection{Calculation of the neutron decay rates}
%%%%%%%%%%%%%%%%%%%%%%%%%%%%%%%%%%%%%%%
\subsubsection{Urca rate}
Traditionally, the Urca rate is decomposed into direct and modified Urca processes.  The direct Urca neutron decay rate is given by the phase space integral
\begin{align}
    \Gamma &= \int \dfrac{\mathop{d^3p_n}}{(2\pi)^3}\dfrac{\mathop{d^3p_p}}{(2\pi)^3}\dfrac{\mathop{d^3p_e}}{(2\pi)^3}\dfrac{\mathop{d^3p_{\bar{\nu}_e}}}{(2\pi)^3}(2\pi)^4\delta^4(p_n-p_p-p_e-p_{\bar{\nu}_e})\nonumber\\
    &\times \dfrac{\sum_{\text{spins}}\vert\mathcal{M}\vert^2}{2^4E_n^*E_p^*E_eE_{\bar{\nu}_e}}f_n(1-f_p)(1-f_e),\label{eq:dUrca_integral}
\end{align}
where $f$ denotes the Fermi-Dirac distribution.  The matrix element, assuming nonrelativistic nucleons and neglecting terms proportional to $(1-g_A^2)$ (see Appendix C in \cite{Harris:2020rus}) is
\begin{equation}
    \dfrac{\sum_{\text{spins}}\vert\mathcal{M}\vert^2}{2^4E_n^*E_p^*E_eE_{\nu}} = 2G_F^2\cos^2{\theta_c}(1+3g_A^2).\label{eq:durca_matrix_element_NR}
\end{equation}
The expression for the direct Urca electron capture rate is identical, except the sign of the neutrino four-momentum is flipped in the delta function and the Fermi-Dirac particle and hole terms are flipped for the neutron, proton, and electron.  These twelve-dimensional integrals can be reduced to three-dimensional numerical integrals, as shown in \cite{Alford:2018lhf,Alford:2021ogv}.

The modified Urca neutron decay and electron capture rates (where the spectator particle is a neutron) are given by eighteen-dimensional phase space integrals similar to Eq.~\ref{eq:dUrca_integral} (see \cite{Alford:2021ogv} for the explicit expressions).  The modified Urca rates can be calculated in the Fermi surface approximation \cite{Alford:2018lhf,Alford:2021ogv}, where it is assumed that the reaction is dominated by the particles near the Fermi surface due to the strongly degenerate nature of the matter at low temperatures.  In this limit, the \textit{net} modified Urca rate (neutron decay minus electron capture) is 
\begin{align}
    &\Gamma_{mU,nd(n)}-\Gamma_{mU,ec(n)} =\nonumber\\
    &\frac{1}{5760\pi^9}G^2g^2_Af^4\frac{(E^*_{Fn})^3E^*_{Fp}}{m^4_{\pi}}\frac{k^4_{Fn}k_{Fp}}{(k^2_{Fn}+m^2_{\pi})^2}\theta_n\label{eq:mUrca_FS}\\
    &\times \delta\mu (1835\pi^6T^6+945\pi^4\delta\mu^2T^4+105\pi^2\delta\mu^4T^2+3\delta\mu^6)\nonumber
\end{align}
where 
\begin{align}
    \theta_n=\begin{cases}
    1 & k_{Fn}>k_{Fp}+k_{Fe}\\
    1-\dfrac{3}{8}\dfrac{(k_{Fp}+k_{Fe}-k_{Fn})^2}{k_{Fp}k_{Fe}} & k_{Fn}<k_{Fp}+k_{Fe}\,.
    \end{cases}
\end{align}
The Fermi surface approximation of the modified Urca rate should break down as the temperature rises and the matter loses its degeneracy.  However, the direct Urca rate alone increases fast enough with increasing temperature that it be taken as infinitely fast (we will see this later) \cite{Alford:2018lhf,Alford:2021ogv} and thus we do not need a better treatment of modified Urca at higher temperatures.  In addition, we mention here that improvements can be made to this rate calculation (see NWA, discussed in next section \cite{Alford:2024xfb}).

When evaluating these rates, we adjust $\mu_{\delta,1}$ to obtain finite-temperature beta equilibrium (c.f.~Eq.~\ref{eq:truebeq_1}).  Note that $\delta\mu$ in  Eq.~\ref{eq:mUrca_FS} is $\mu_n-\mu_p-\mu_e$, so in finite-temperature beta equilibrium, the net modified Urca rate is not necessarily zero.  An example of how both direct Urca processes and the modified Urca processes change when $\mu_{\delta,1}=0$ or in finite-temperature beta equilibrium is given in Figs.~6 and 7 of \cite{Alford:2018lhf}.  To determine $\lambda_1$, we calculate $\Gamma_{n\rightarrow p}-\Gamma_{p\rightarrow n}$ as a function of $\delta\mu_1$ around finite-temperature beta equilibrium, where $\Gamma_{n\rightarrow p}-\Gamma_{p\rightarrow n}$ crosses through zero: the slope of that line is $\lambda_1$ \cite{Alford:2018lhf,Alford:2021ogv}.

\begin{figure*}\centering
\includegraphics[width=0.45\textwidth]{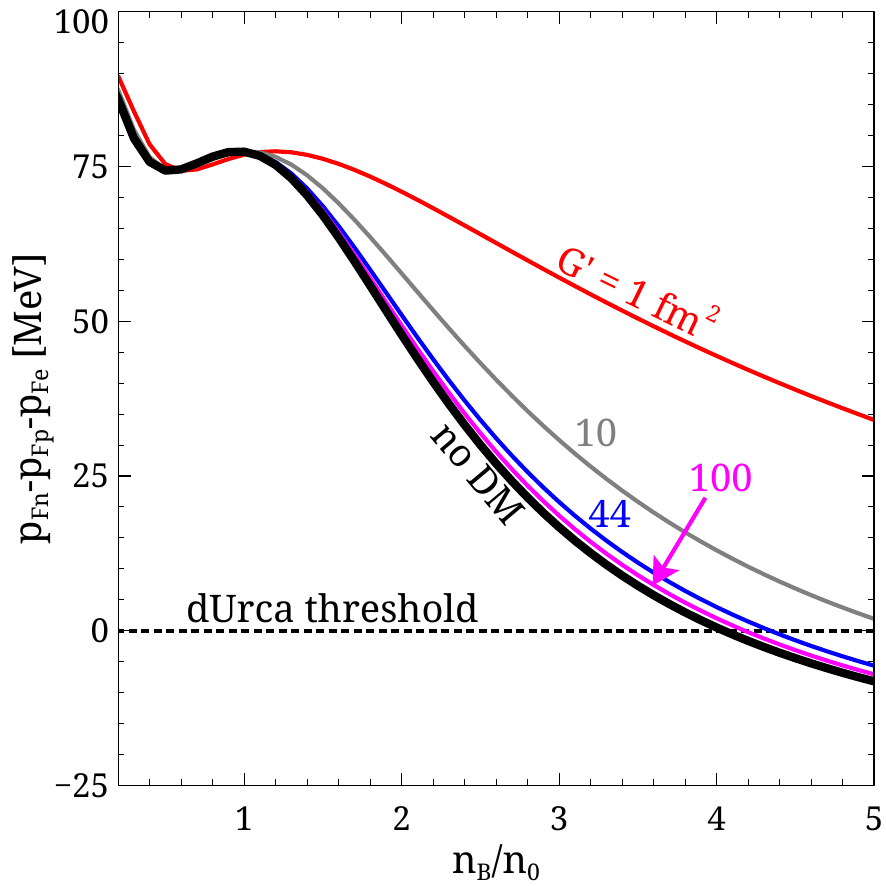}\hspace{1cm}
\includegraphics[width=0.45\textwidth]{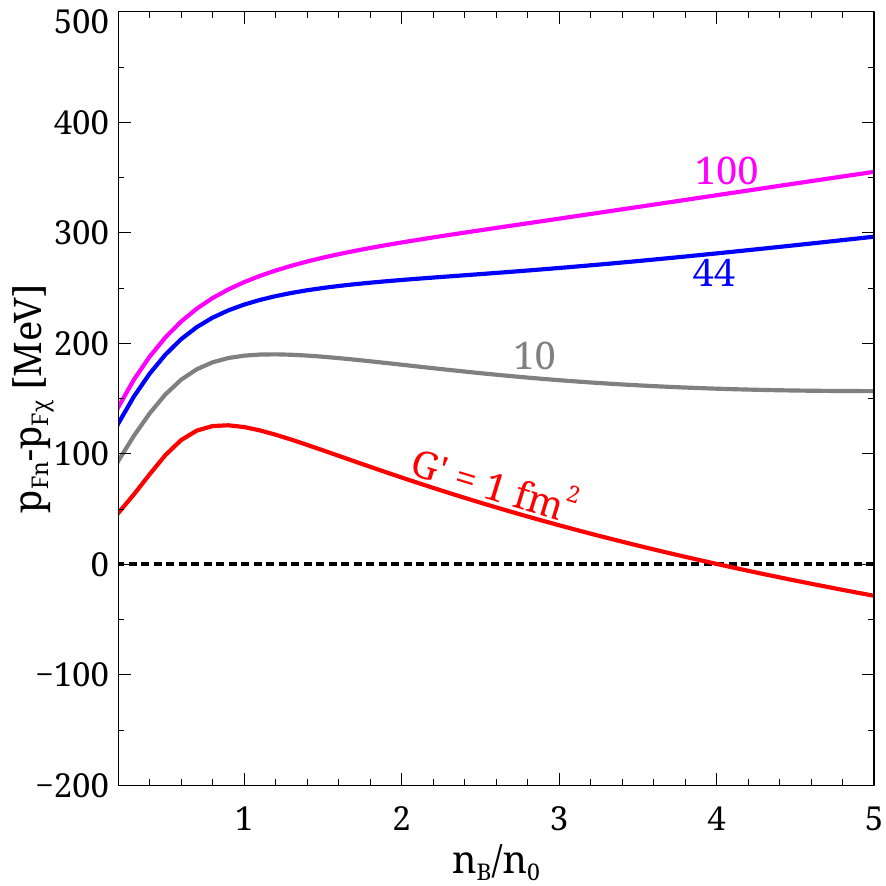}
\caption{Kinematic suppression factors of neutron decays.  Note the different y-axis scales.  The calculations in these plots are done at zero temperature and the zero-temperature beta equilibrium conditions are used.}
\label{fig:kinematic_suppression}
\end{figure*}

Perhaps the key feature of the Urca process at low temperature is the presence of the direct Urca threshold (density).  Around nuclear saturation density, $npe$ matter is extremely neutron rich, with a proton fraction around 0.05.  We see this for the IUF-II EoS in Fig.~\ref{fig:particlefractions}, but most models of dense matter agree on this \cite{Keller:2022crb,Tsang:2023vhh}, due to our improving knowledge of the symmetry energy at these relatively low densities \cite{Li:2019xxz,Koehn:2024set,Lattimer:2023rpe}.  With such a small proton (and electron) population, it is impossible for neutrons, protons, and electrons on their respective Fermi surfaces to conserve momentum, and thus the direct Urca rate is Boltzmann suppressed, as particles away from their Fermi surfaces dominate the rates \cite{Alford:2018lhf,Alford:2021ogv}.  As density increases, the symmetry energy and proton fraction grow and when the proton fraction exceeds 1/9, $p_{Fp}+p_{Fe}>p_{Fn}$ and direct Urca becomes kinematically allowed.  As temperature rises, particles away from the Fermi surface can participate in reactions and the direct Urca threshold becomes strongly blurred and, eventually, essentially meaningless.  

An interesting question is whether the presence of the dark baryons changes the direct Urca threshold.  In the left panel of Fig.~\ref{fig:kinematic_suppression}, we plot the Fermi momentum difference $p_{Fn}-p_{Fp}-p_{Fe}$ versus density.  When the Fermi momentum difference goes negative, direct Urca becomes allowed.  The direct Urca threshold for IUF-II without dark baryons is around $4n_0$.  Adding a small number of dark baryons (by choosing a large repulsion strength $G'$) leads to a small change in the direct Urca threshold, as (c.f.~Fig.~\ref{fig:particlefractions}) the neutron, proton, and electron populations all decrease slightly to accommodate the presence of the dark baryons.  Evidently, this slightly worsens the momentum deficit and the threshold is pushed to slightly higher density.  The change, however, is well within the uncertainty in the $npe$ EoS direct Urca threshold, which is quite large still \cite{Tsang:2023vhh}.  When the repulsion strength is decreased further, beyond what is allowed by a $2M_{\odot}$ neutron star, the dark matter population becomes substantial and direct Urca becomes increasingly kinematically forbidden.  
%%%%%%%%%%%%%%%%%%%%%%%%%%%%%%%%%%%%%%%%%%%
\subsubsection{Neutron dark decay rate}\label{sec:neutron_dark_decay}
The rate of the ``direct'' neutron dark decay process $n\rightarrow\chi\phi$ is given by the phase space integral 
\begin{align}
    \Gamma_{n\rightarrow\chi\phi}^{\text{direct}} &= \int \dfrac{\mathop{d^3p_n}}{(2\pi)^3}\dfrac{\mathop{d^3p_{\chi}}}{(2\pi)^3}\dfrac{\mathop{d^3p_{\phi}}}{(2\pi)^3}(2\pi)^4\delta^4(p_n-p_{\chi}-p_{\phi})\nonumber\\
    &\times \dfrac{\sum_{\text{spins}}\vert\mathcal{M}\vert^2}{2^3E_n^*E_{\chi}^*E_{\phi}}f_n(1-f_{\chi}),\label{eq:dUrca_darkdecay_integral}
\end{align}
where the spin-summed matrix element is
\begin{equation}
    \sum_{\text{spins}}\vert\mathcal{M}\vert^2 = 4g_{\phi}^2\left(\tilde{p}_n\cdot\tilde{p}_{\chi}+m_*m_{\chi}\right).
\end{equation}
There is no Bose factor for the $\phi$, because the $\phi$ particles are assumed to escape the star.  This matrix element is different from the vacuum one (Eq.~\ref{eq:nchiphi_matrix_element}) because in deriving the vacuum matrix element we used energy-momentum conservation relations in a form which applies in vacuum, but not in medium (due to the mean-field contributions to the energy of the neutron and dark baryon in an RMF theory \cite{Roberts:2016mwj}).  Here, we have to stick with the form given above.  The momenta four-vectors $\tilde{p}$ are $\left(E^*,\mathbf{p}\right)$ where $E^*$ is the energy without the vector mean field contribution $U$: $E^*\equiv E-U$ (see Appendix B of \cite{Roberts:2016mwj}).  This integral can be simplified into a two-dimensional numerical integral.  The details are given in the Appendix \ref{sec:appendix_ndecay_direct}.  The backwards rate $\chi\rightarrow n+\phi$ is the same phase space integral as Eq.~\ref{eq:dUrca_darkdecay_integral} but with the sign of $p_{\phi}$ reversed in the delta function and with the replacements $f_n\rightarrow (1-f_n)$ and $(1-f_{\chi})\rightarrow f_{\chi}$.

The usual approach would be, following the Urca example, to calculate the rate of the ``modified'' dark decay processes, such as $n+n\rightarrow n+\chi+\phi$ and $n+\chi \rightarrow \chi + \chi+\phi$ and their inverses.  However, such a high-dimensional phase space integral (15, in this case) is difficult to do numerically\footnote{In fact, attempting to do this integral ``exactly'' fails, as for some momentum combinations in the integral, the neutron and/or dark baryon propagators go on shell, and the integral diverges.  This happens in the modified Urca case too.  In the degenerate limit, the momenta all take fixed values, so there is no risk of the propagator diverging.  Further details are given in \cite{Alford:2024xfb,Shternin:2018dcn,Suleiman:2023bdf}.}, so the typical calculation method is to calculate it in the degenerate limit, where one assumes all of the participating fermions are on their Fermi surface.  For example, this is how the modified Urca rate is calculated \cite{Shapiro:1983du,Yakovlev:2000jp}.  For cold neutron stars, this is sufficient\footnote{However, the conventional rate of modified Urca invokes other approximations that lead to an error which is up to an order of magnitude \cite{Shternin:2018dcn,Alford:2024xfb}.}, but as we will see soon, we need to be able to calculate the rate of $n+n\rightarrow n+\chi+\phi$ at temperatures of several tens of MeV, where the matter is certainly not strongly degenerate.  However, in Appendix \ref{sec:appendix_ndecay_modified1} and \ref{sec:appendix_ndecay_modified2} we detail the Fermi surface approximation calculation of the ``modified'' neutron dark decay processes.  To calculate the neutron dark decay rate at high temperature, we must use a more advanced, recently developed method called the nucleon width approximation (NWA) \cite{Alford:2024xfb}.  We do not use the NWA approximation for the Urca rate because we do not need to know its precise value at temperatures of tens of MeV (one can treat the rate as being infinitely fast) and also because the influence of the NWA method on the Urca bulk viscosity (at temperatures of a few MeV) deserves its own dedicated study.   

The decomposition of a rate, e.g.~Urca, into direct and modified contributions is an attempt to account for the collisional broadening of strongly interacting particles in dense matter.  In principle, this expansion should include the possibility of two or more spectator particles as well.  NWA incorporates the effects of these collisions into an imaginary part added to the nucleon mass \cite{Alford:2024xfb}.  The imaginary part is $iW_i/2$ proportional to the nucleon width $W_i$ which is a function of density and temperature.  The propagator of a nucleon with a finite width can be written as an integral over mass of the zero-width propagators, weighted by a Breit-Wigner factor with width $W_i$ - called the mass-spectral decomposition \cite{Kuksa:2015iaa} - which leads to the following formula for calculating rates within the NWA approximation.  If the rate of the direct $n\rightarrow\chi+\phi$ process is $\Gamma_{n\rightarrow\chi\phi}^{\text{direct}}$ (which is a function of, among other things, $m_*$ and $m_{\chi}$) and we assume that the neutron and $\chi$ have widths $W_n$ and $W_{\chi}$, respectively, (the $\phi$ is assumed not to have a width), then the rate of that process in the nucleon width approximation is
\begin{equation}
    \Gamma^{\text{NWA}}_{n\rightarrow\chi\phi} = \int_{-\infty}^{\infty}\mathop{dm_*}\mathop{dm_{\chi}}\Gamma_{n\rightarrow\chi\phi}^{\text{direct}}R_nR_{\chi},\label{eq:NWA_massmassintegral}
\end{equation}
where $R_n$ and $R_{\chi}$ are the Breit-Wigner factors
\begin{subequations}
\begin{align}
    R_n &= \dfrac{1}{2\pi}\dfrac{W_n}{(m_*-m_*^{\text{EOS}})^2+(W_n/2)^2}\\
    R_{\chi} &= \dfrac{1}{2\pi}\dfrac{W_{\chi}}{(m_{\chi}-m_{\chi}^{\text{vac}})^2+(W_{\chi}/2)^2}.
\end{align}
\end{subequations}
Here, $m_*^{\text{EOS}}$ means the EoS-specified value of the nucleon Dirac effective mass, which is a function of density and temperature, and $m_{\chi}^{\text{vac}}$ is the vacuum mass of the $\chi$ particle, here 938 MeV, but 940 MeV in Sec.~\ref{sec:rapid_n_decays}.  In other words, the NWA rate takes the traditional calculation of the $n\rightarrow\chi+\phi$ rate $\Gamma_{n\rightarrow\chi\phi}^{\text{direct}}$ and evaluates it at all possible neutron effective masses and $\chi$ masses, but weighted with a Breit-Wigner factor around the actual neutron effective mass (at the given density and temperature) and the vacuum value of the $\chi$ mass. 

For the widths, in the original NWA paper Alford \textit{et al.}~\cite{Alford:2024xfb} used $W\approx T^2/(5\text{ MeV})$ which comes from the calculation in \cite{Sedrakian:2000kc}.  Here, we need both the neutron and $\chi$ widths, and the $\chi$ width will be a function of the parameter $G'$.  To calculate the neutron width, we calculate the inverse mean free path of the neutron due to the elastic scattering process $n+n\rightarrow n+n$.  We assume that the neutrons exchange either a sigma or an omega meson, and we include the interference terms.  For the $\chi$, we considered $\chi+\chi\rightarrow \chi+\chi$ scattering via a dark vector exchange (the same consideration used in obtaining the cross section Eq.~\ref{eq:sigma_chichi}).  As with $G'$, we define $G_i \equiv \left(g_i/m_i\right)^2$.  In the degenerate limit, we find the widths (evaluated for a particle on the Fermi surface) to be
\begin{subequations}
\begin{align}
W_n &= \dfrac{1}{480\pi}\dfrac{T^2}{E_{Fn}^*}\bigg[15m_*^4\left(G_{\sigma}-G_{\omega}\right)^2\label{eq:WN_degenerate}\\
&+10\left(G_{\sigma}^2-5G_{\sigma}G_{\omega}+4G_{\omega}^2\right)p_{Fn}^2m_*^2\nonumber\\
&+\left(7G_{\sigma}^2+12G_{\sigma}G_{\omega}+76G_{\omega}^2\right)p_{Fn}^4\bigg]\nonumber\\
W_{\chi} &= \dfrac{1}{480\pi}\dfrac{T^2}{E_{F\chi}^*}G'^2\left[15m_{\chi}^4+40p_{F\chi}^2m_{\chi}^2+76p_{F\chi}^4\right],\label{eq:Wchi_degenerate}
\end{align}
\end{subequations}
In the above formulas, $E_{F,i}^*$ is the Landau effective mass of the particle species (in contrast to the Dirac effective mass of the neutron $m_*$).  The values of $G_i$ for the $npe$ matter EoSs used in this work are shown in Table \ref{table:Gvalues}.  In our rate calculations, we do the full phase space integration with arbitrary particle degeneracy, but showing the widths in the degenerate limit allows us to see some characteristic features.  For example, the widths indeed go as the square of the temperature, as they should \cite{coleman2015introduction}.  Secondly, the nucleon width has, in the limit of low density (small Fermi momentum), a cancellation between the scalar and vector exchange terms.  No such cancellation exists for the dark baryon, because we only gave it repulsive (vector) interactions, and so the nucleon width will typically be smaller than the dark baryon width in our model.

We should also note that the widths are a function of the energy of the particle.  In the width calculation, we consider an on-shell particle, but of course one should be more general in the future.  In the degenerate limit expressions (Eqs.~\ref{eq:WN_degenerate} and \ref{eq:Wchi_degenerate}), we chose the energy to be the Fermi energy, but in the full-phase-space-integration approach (valid for arbitrary degeneracy), which we use in our actual rate calculations, we chose a typical energy: for example, for the nucleons,
\begin{equation}
    E_n^{*,\text{avg.}} = \dfrac{\int\dfrac{\mathop{d^3k}}{\left(2\pi\right)^3}\sqrt{k^2+m_*^2}f_n\left(1-f_n\right)}{\int\dfrac{\mathop{d^3k}}{\left(2\pi\right)^3}f_n\left(1-f_n\right)}.
\end{equation}
and similar for the dark baryons.  This choice is motivated by the ``effective density" defined in some earlier works (e.g.~the appendix of \cite{Raffelt:1996wa} and also \cite{Payez:2014xsa}).

\begin{figure*}\centering
\includegraphics[width=0.45\textwidth]{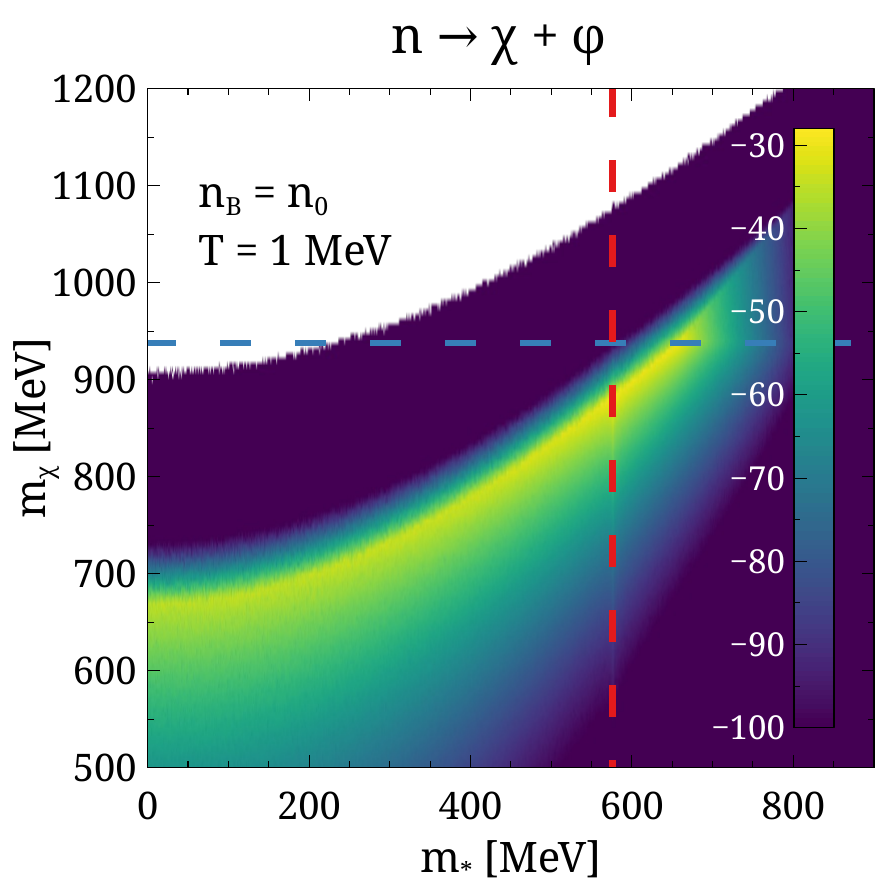}\hspace{1cm}
\includegraphics[width=0.45\textwidth]{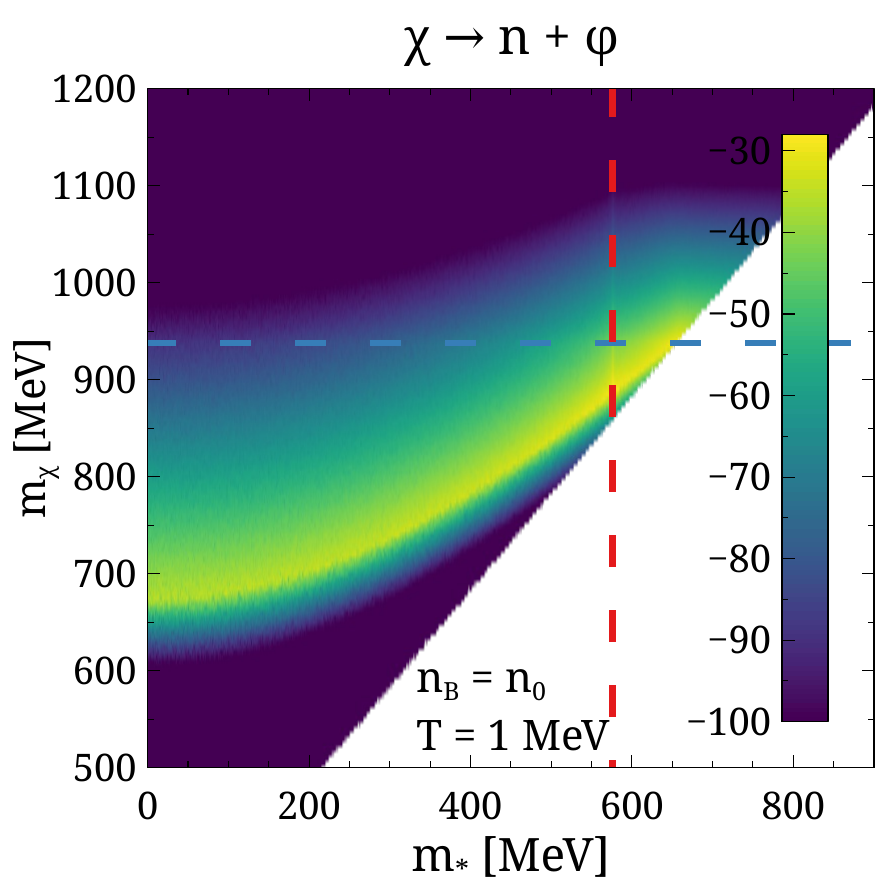}
\caption{The NWA rate integrand $\Gamma_{n\rightarrow\chi\phi}^{\text{direct}}R_nR_{\chi}$ (c.f.~Eq.~\ref{eq:NWA_massmassintegral}) plotted with a log scale in the $m_*m_{\chi}$ plane, for $n\rightarrow\chi+\phi$ (left panel) and $\chi\rightarrow n+\phi$ (right panel), at $n_B=n_0$ and $T=1\text{ MeV}$.  Dotted lines indicate the nucleon effective mass $m_*^{\text{EoS}}$ at $n_B=n_0$ and $T=1\text{ MeV}$, as given by the EoS, and the vacuum $\chi$ mass $m_{\chi}^{\text{vac}}$.  In these conditions, the neutron width is 100 keV and the $\chi$ width is 11 MeV.  The color bar scale shows $\log_{10}\left(\Gamma_{n\rightarrow\chi\phi}^{\text{direct}}R_nR_{\chi}/\text{MeV}^2\right)$.}
\label{fig:NWA_massmassplane}
\end{figure*}

In Fig.~\ref{fig:NWA_massmassplane}, we plot the integrand of the NWA integral, for the forward and reverse processes.  The NWA integral takes place at a fixed density and temperature, where the chemical potentials $\mu_n^*$ and $\mu_{\chi}^*$ are fixed.  But, the masses of the particles $m_*$ and $m_{\chi}$ are integrated over, and thus in the phase space integral, the Fermi momenta of the particles vary according to $k_{Fn}=\sqrt{(\mu_n^*)^2-m_*^2}$ and $k_{F\chi}=\sqrt{(\mu_{\chi}^*)^2-m_{\chi}^2}$ (of course, this is only a meaningful statement when the NWA integral is evaluated at high density and low temperature).  The NWA integrand is peaked (indicated by the yellow region in Fig.~\ref{fig:NWA_massmassplane}) for the mass combinations which yield neutron and $\chi$ Fermi momenta which are nearly equal.  Just as the direct Urca threshold is given by $p_{Fn}=p_{Fp}+p_{Fe}$, the neutron dark decay ``threshold'' is $p_{Fn}=p_{F\chi}$.  However, while for direct Urca, one can be ``above threshold'', just by putting the proton and electron momenta at an angle with each other, in the neutron dark decay process, one can never be above threshold because there are two degenerate particle species, not three (c.f.~Eq.~12 in \cite{Kaminker:2016ayg}).  A final feature of the NWA plot is that demanding that the $\phi$ energy be positive means that for $n\rightarrow\chi+\phi$, the neutron energy must be larger than the $\chi$ energy, and thus the $\chi$ mass must not be too large, while for $\chi\rightarrow n+\phi$, the $\chi$ energy must be larger than the neutron energy and thus the neutron effective mass must not be too large.  These features are visible in Fig.~\ref{fig:NWA_massmassplane}.

\begin{figure}\centering
\includegraphics[width=0.45\textwidth]{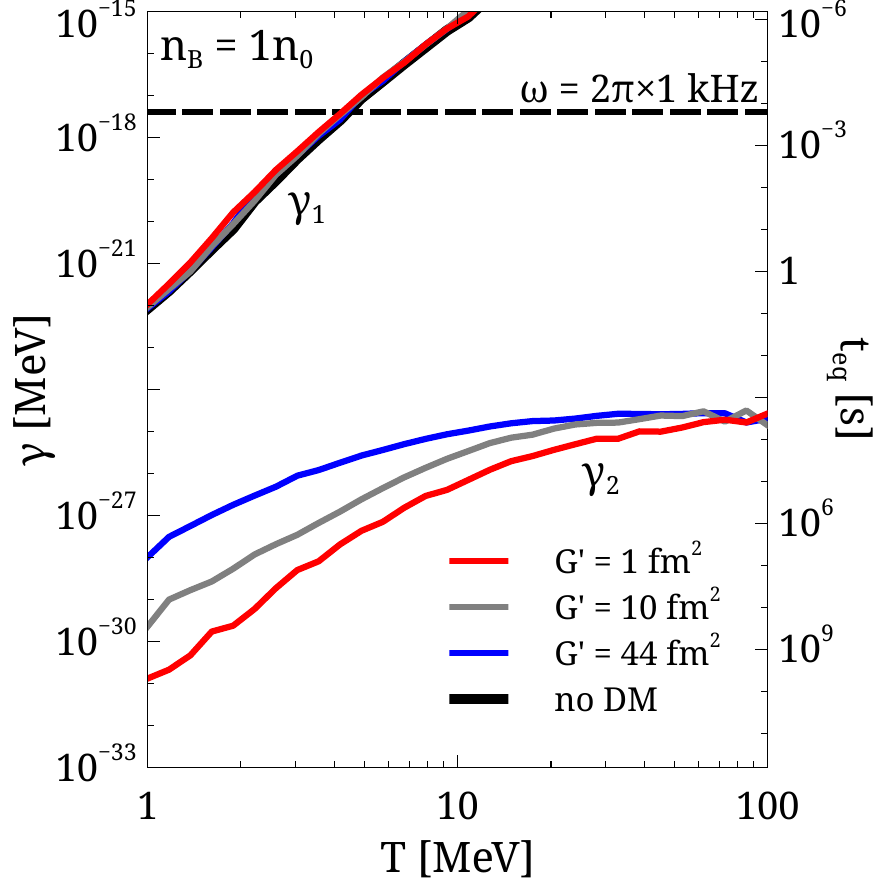}
\caption{Beta equilibration rates $\gamma_1$ and $\gamma_2$, calculated at $n_B=n_0$, as a function of temperature.  The coupling $g_{\phi}$ ($\gamma_2\sim g_{\phi}^2$) is set so that the neutron dark decay has a 1\% branching ratio.  Different self-interaction strengths for the dark baryons are shown in different colors.  For $\gamma_1$, the results differ by, at most, a factor of two, so the curves are essentially overlapping.  The right vertical axis shows the beta-equilibration timescale $t_{\text{eq},i}\equiv 1/\gamma_i$.}
\label{fig:gammas}
\end{figure}

In Fig.~\ref{fig:gammas}, we plot the beta equilibration rates $\gamma_1$ and $\gamma_2$ (Eqs.~\ref{eq:gamma1} and \ref{eq:gamma2}) as a function of temperature.  Beta equilibration proceeds as $\exp{\left(-\gamma t\right)}$ and thus the timescale of equilibration is $\gamma^{-1}$, which is shown on the right vertical axis of Fig.~\ref{fig:gammas}.  Different choices of dark baryon self-repulsion are shown in different colors, and the pure $npe$ matter case is shown in solid black.  They are calculated using the finite-temperature equilibrium conditions Eqs.~\ref{eq:truebeq_1} and \ref{eq:truebeq_2}.  As with $\lambda_1$, we found $\lambda_2$ by calculating $\Gamma_{n\rightarrow \chi}-\Gamma_{\chi\rightarrow n}$ as a function of $\delta\mu_2$ around finite-temperature beta equilibrium: the slope of that line is $\lambda_2$.  

Fig.~\ref{fig:gammas} shows that the Urca rate rises monotonically with temperature, reaching resonance (1 kHz) at temperatures of 4-5 MeV.  Including dark baryons in the EoS has little effect on the Urca rate.  This was evident in Fig.~\ref{fig:bulk_viscosity_l2zero}, as all bulk viscosity curves peaked at nearly the same temperatures: 4-5 MeV.  

The dark baryon equilibration rate $\gamma_2$ depends on the self-interaction strength as would be expected, as the dark baryon width is a function of $G'$ and $G'$ also influences the $\chi$ population.  The rate $\gamma_2$ rises with temperature, but then levels off as the temperature surpasses 10 MeV.  This is likely because at a few tens of MeV, both the neutron and the dark baryon are nondegenerate, and so not much phase space is gained as temperature increases.  In contrast, in the Urca process, the electron is still degenerate at temperatures of a few tens of MeV, and thus its available phase space is still increasing as temperature rises.  

In medium, the neutron dark beta equilibration rate is much more than several hundred times slower than the Urca rate, which was the benchmark ratio in vacuum.  Indeed, the dark baryon beta equilibration rate in medium is slower than all other scales in the problem: even at high temperatures its equilibration timescale is about 40 minutes, and thus we were justified in setting $\gamma_2=0$ in producing Fig.~\ref{fig:bulk_viscosity_l2zero}.

In part, this vast difference is because the Urca rate itself benefits from the large electron phase space ($\sim p_{Fe}^2T$), which has no counterpart in the neutron dark decay.  In addition, part of the reason why the neutron dark decay rate is so slow is shown in the right panel of Fig.~\ref{fig:kinematic_suppression}.  In degenerate conditions where particles near the Fermi surface are the only ones that can participate in a reaction, the rate $n\rightarrow\chi+\phi$ is only unsuppressed when the Fermi momenta of the neutron and dark baryon are equal (the beta equilibrium condition guarantees their Fermi energies are equal).  But, for a reasonable value of $G'$, the $\chi$ population is much smaller than the neutron population, guaranteeing that the Fermi momenta are mismatched (by one hundred or more MeV!).  In fact, Fig.~\ref{fig:kinematic_suppression} shows that as density increases, this problem gets even worse, which is the opposite of the behavior of the Urca process.  So, the neutron dark decay is even more kinematically suppressed than is Urca, in addition to having a small coupling constant $g_{\phi}$, which multiplicatively suppresses the rate.

Changing the dark baryon repulsion strength $G'$ has two effects.  First, increasing $G'$ leads to fewer dark baryons present in the system.  This leads to more extreme kinematic suppression of the neutron dark decay process (Fig.~\ref{fig:kinematic_suppression}).  However, the increased interaction strength between $\chi$ particles actually enhances the rate of spectator processes like $n+\chi\rightarrow \chi+\chi+\phi$ - apparently enough to, on net, increase the rate of beta equilibration (Fig.~\ref{fig:gammas}).  

Thus, if the decay $n\rightarrow\chi+\phi$ can proceed in vacuum (and thus has a branching ratio well under 1\%), and if the dark $\phi$ particles are low-mass and are not trapped in the merger, then the dark baryon population is essentially frozen on the timescales of a neutron star merger, except for a very low-mass merger that produces a stable, or very long-lived remnant.  Bulk viscosity in this case is due to just the Urca process, and the dark baryons only have a slight impact on the EoS and the bulk viscosity.  
%%%%%%%%%%%%%%%%%%%%%%%%%%%%%%%%%%%%%%%%%%%%%%%%
\subsection{Bulk viscosity with rapid production of dark baryons}\label{sec:rapid_n_decays}
We introduced a dark baryon $\chi$ that was lighter than the neutron, because the model was originally motivated by the neutron decay anomaly.  However, dark sectors coupled to neutrons are of general interest.  If the dark baryon $\chi$ is heavier than the neutron, then the neutron lifetime in vacuum is in no way altered and $g_{\phi}$ becomes a free parameter (though still limited by the Raffelt criterion from above (see below) and cold neutron stars from below).  We fix the value $m_{\chi} = 940\text{ MeV}$ in this section and keep the $\phi$ massless for convenience.  We also will keep $G'=44\text{ fm}^2$.

The Raffelt criterion demands that the $\phi$ luminosity from a supernova environment must be less than about $2\times 10^{52}\text{ ergs/s}$ \cite{Raffelt:1996wa,Raffelt:1990yz}, or else the neutrino signal from SN1987a would be different from what was observed.  Integrating the $\phi$ emissivity over a simple analytic supernovae profile \cite{Chang:2016ntp}, one finds that this corresponds to
\begin{equation}
    g_{\phi} \lesssim 2\times 10^{-10} \quad \text{Raffelt limit}.
\end{equation}
The choice of profile could change this bound by a factor of a few \cite{Chang:2016ntp}.  Of course, if $g_{\phi}$ is large enough, the $\phi$ particle becomes trapped, and different methods \cite{Caputo:2022mah,Fiorillo:2025yzf,Dev:2021kje,Horowitz:2012jd,Fiorillo:2025sln} must be used to constrain it.  

\begin{figure*}\centering
\includegraphics[width=0.45\textwidth]{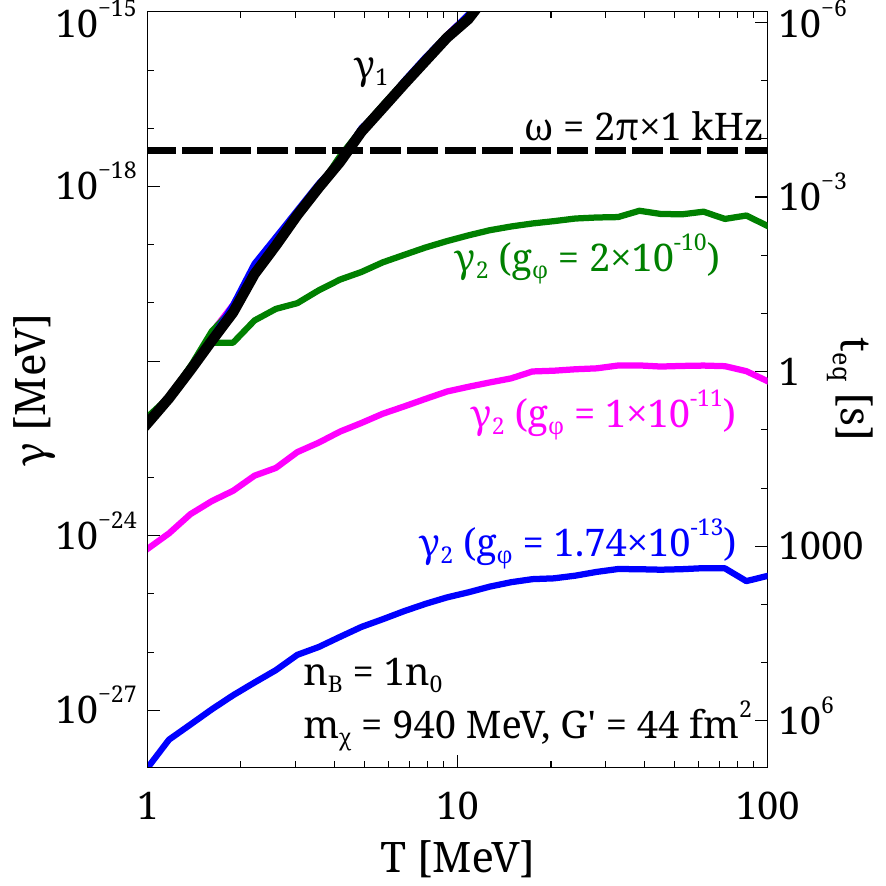}\hspace{1cm}
\includegraphics[width=0.45\textwidth]{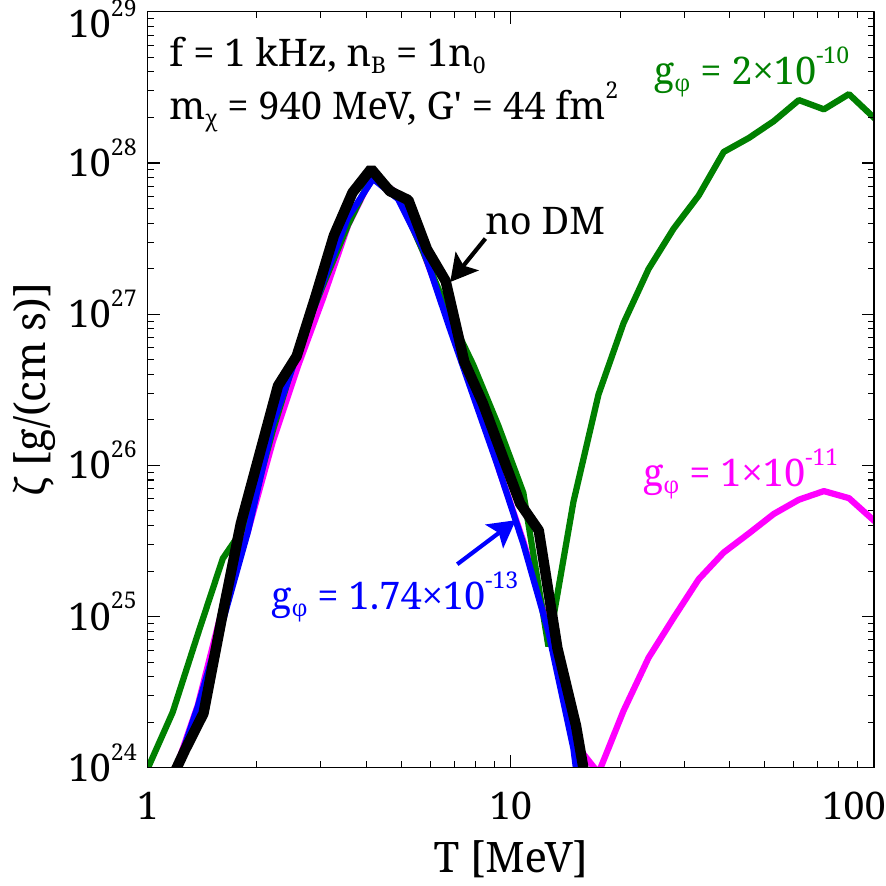}
\caption{Left: Beta equilibration rates $\gamma_1$ and $\gamma_2$, calculated at $n_B=n_0$, as a function of temperature.  Different color curves correspond to different choices of $g_{\phi}$  ($\gamma_2\sim g_{\phi}^2$).  The right vertical axis shows the beta-equilibration timescale $t_{\text{eq},i}\equiv 1/\gamma_i$.  Right: Bulk viscosity of the $npe\chi$ matter in response to a 1 kHz density oscillation, for each of the choices of $g_{\phi}$ made in the left panel.  In these figures, the dark baryon has a $m_{\chi}=940\text{ MeV}$ so that it does not influence the neutron lifetime (in vacuum) at all.  The coupling $g_{\phi}$ is set to arbitrary values.}
\label{fig:fastdarkdecay}
\end{figure*}

We choose several values for the coupling $g_{\phi}$ which range from the one that yields a 1\% branching ratio to one that is about three orders of magnitude higher and saturates the Raffelt bound.  Our goal is to see if allowing for this faster neutron decay can lead to changes in the bulk viscosity.  The equilibration rate $\gamma_2$ goes as $g_{\phi}^2$, so a three order of magnitude increase in $g_{\phi}$ leads to a six order of magnitude increase in the rate $\gamma_2$, which can be seen in the left panel of Fig.~\ref{fig:fastdarkdecay}.  Even at the maximum allowed value of $g_{\phi}$, the neutron dark decay is slower than the 1 kHz density oscillation.  However, it is much closer to resonance than before.  Notably, at $T\approx 1\text{ MeV}$, the neutron decay rate via Urca and via dark decay occur at nearly equal rates when $g_{\phi}$ is at the Raffelt limit.  

In the right panel of Fig.~\ref{fig:fastdarkdecay}, we plot the bulk viscosity of $npe\chi$ matter with the three choices of $g_{\phi}$ shown in the left panel.  As before, the pure $npe$ matter and the 1\% branching ratio $npe\chi$ matter bulk viscosities are quite close, displaying just the Urca bulk-viscous peak at $T=4-5\text{ MeV}$, when $g_{\phi}$ is increased to $10^{-11}$ or the Raffelt bound, $2\times 10^{-10}$, the bulk viscosity is strongly enhanced at temperatures of many tens of MeV.  This is due to the dark baryon equilibration rate approaching resonance as temperature increases.  We note that while in the right panel of Fig.~\ref{fig:fastdarkdecay}, the pink and green curves (the two with the highest value of $g_{\phi}$) appear to exhibit a peak, this is different from the Urca peak at lower temperature because in the case of Urca, the beta equilibration rate $\gamma_1$ increases monotonically with temperature, passing through the resonant value of 1 kHz, while the dark rate as a function of temperature reaches a maximum and then decreases slightly, never passing through the value of 1 kHz.

\begin{figure}\centering
\includegraphics[width=0.45\textwidth]{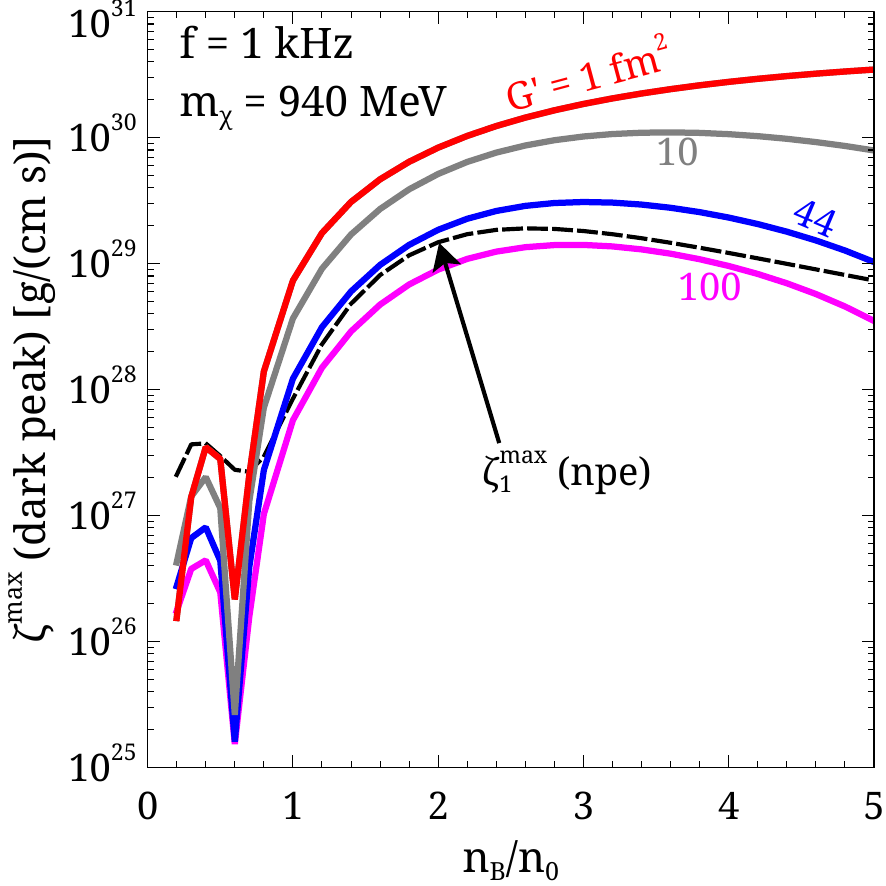}
\caption{Peak value (across all temperatures) of the dark decay bulk viscosity $\zeta_2$, as a function of the baryon density.  A variety of dark baryon self-repulsion strengths $G'$ are displayed.  Also displayed, in dashed black, is the peak value of the Urca bulk viscosity in the pure $npe$ matter case, for comparison.}
\label{fig:zeta2_max}
\end{figure}

We can calculate the value of the maximum value of the partial bulk viscosity $\zeta_2$ in an effort to determine how large the dark bulk viscosity could actually be if resonance were reached (which we reiterate, does not occur here because $g_{\phi}$ is limited by the Raffelt bound in this model).  This $\zeta_2^{\text{max}}$ quantity is plotted in Fig.~\ref{fig:zeta2_max}.  As the peak of $\zeta_2$ is larger than the peak value of $\zeta_1$, the difference in adiabatic and equilibrium (with respect to the dark baryon number) sound speeds is greater than adiabatic and equilibrium (with respect to the proton fraction) sound speeds, allowing the peak value of $\zeta_2$ to exceed $\zeta_1$ (c.f.~Eq.~\ref{eq:A1sqB1} and \ref{eq:A2sqC2}).  However, the actual value of bulk viscosity achieved by Raffelt-criterion-consistent dark decays is a factor of 10 less than the peak value.  

\begin{figure}[t!]\centering
\includegraphics[width=0.45\textwidth]{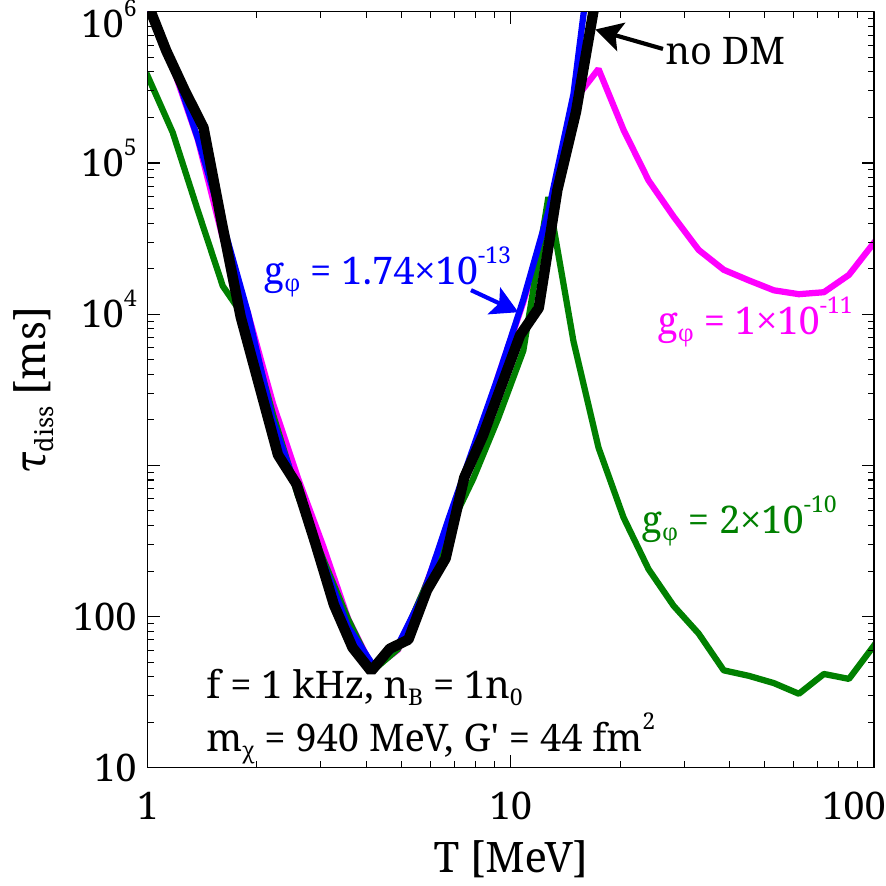}
\caption{Bulk-viscous dissipation timescale, for a 1 kHz density oscillation, as a function of temperature.  The density is $n_B=n_0$.  The colors correspond to the values of $g_{\phi}$ chosen in Fig.~\ref{fig:fastdarkdecay}.}
\label{fig:tdiss_fast}
\end{figure}

Finally, we calculate the timescale on which bulk viscosity damps a density oscillation \cite{Alford:2019qtm}
\begin{equation}
    \tau_{\text{diss}}=\dfrac{\varepsilon}{\mathop{d\varepsilon}/\mathop{dt}} = \dfrac{\kappa_T^{-1}}{\omega^2\zeta}.
\end{equation}
As defined in Eq.~\ref{eq:compressibility}, $\kappa_T$ is the compressibility of dense matter, and thus its inverse is the incompressibility, which acts like the spring constant in the energy of a density oscillation in dense matter.  We plot the result in Fig.~\ref{fig:tdiss_fast}.  Essentially, this figure is the inverse of the bulk viscosity.  The Urca bulk viscosity damps density oscillations (at saturation density) in as little as 40 ms, which is roughly consistent with previous estimates, and without the influence of the neutron dark decay rate, the damping time rises rapidly at higher temperatures, and bulk viscosity becomes essentially unable to damp oscillations out in any reasonably timescale.  However, neutron dark decays allow for rapid damping of oscillations in matter of many tens of MeV temperatures, perhaps as quickly as 20 ms timescales.     
%%%%%%%%%%%%%%%%%%%%%%%%%%%%%%%%%%%%%%%%%
\section{Conclusions}
In this work, we investigated the possible impacts of a novel neutron decay $n\rightarrow\chi+\phi$ on transport in dense matter, in particular, the bulk viscosity of matter in the hot, dense environment of a neutron star merger.  We significantly improved the calculation of the decay rate in medium compared to previous estimates, using the nuclear width approximation (NWA) method to include the effects of spectator particles on the decay rate.  We found that if the dark sector parameters are such that the neutron decays (in vacuum) in the dark channel 1\% of the time, which was thought to explain the neutron lifetime anomaly, then in medium, due to the large difference between the neutron and $\chi$ densities, the neutron dark decay rate is quite slow compared to the dynamical timescale of a neutron star merger.  Furthermore, the effect of the dark baryons on the EoS decreases the Urca bulk viscosity slightly (a factor of 2-3 at most), but the change is well within the current uncertainty in the $npe$-matter EoS.    

We then considered the same model, but where the dark baryon is heavier than the neutron, so that the neutron lifetime in vacuum is undisturbed.  In this situation, the $n\chi\phi$ coupling can be much larger, and the neutron dark decay rate in medium can be significantly faster.  At high temperatures, it becomes comparable to other timescales in the problem, and in fact leads to a large bulk viscosity at temperatures of many tens of MeV.  The specific model we study predicts density oscillations at, say, $T = 50\text{ MeV}$, damp out due to bulk viscous dissipation in under 40 ms.  

These observations lead to a general point: typical standard model reactions are relatively fast.  The slow, ``weak'' interactions like the Urca process are still fast enough to reach millisecond timescales at temperatures of only a few MeV.  At the high temperatures that exist in much of a neutron star merger remnant or a supernovae, all rates are fast enough to keep matter in chemical equilibrium no matter the degree of freedom.  However, BSM physics likely has very weak couplings, or else it would have been found already in laboratory experiments.  These couplings may be small enough to keep the rates sufficiently slow so that only in high temperature environments do they become comparable to the dynamical timescale of the system.  Maybe a signature of dark sectors is significant bulk-viscous dissipation in environments with temperatures of many tens of MeV.  

Even if the neutron dark decay is slow compared to the millisecond timescales relevant to bulk-viscous dissipation in neutron star merger remnants, it might well have observable consequences in, say, protoneutron star environments.  For example, the release of gravitational binding energy as neutrons in the protoneutron star decay into dark baryons could produce a long-duration neutrino signal.

In this work, we assumed the dark baryons were in thermal equilibrium with the $npe$ matter in the neutron star environment.  However, if the dark baryons are dark matter, their interaction with nucleons is quite feeble, and therefore it is likely that a two-fluid approach is more realistic.  Bulk viscosity in this two-fluid context should be calculated.  Even in the one-fluid perspective, further investigations are warranted.  In this work, we considered a dark baryon that was much heavier than the dark scalar, but the reverse case, or near-equal masses, could be examined.  Also, it is possible that the $\phi$ particles could be trapped in the system, due to some unspecified reaction or due to very large values of $g_{\phi}$.  In this case, reactions like $n\rightarrow\chi+\phi$ and $\chi\rightarrow n+\phi$ can proceed forward and backward, which could well have interesting effects on the bulk viscosity.  Other possible neutron dark decays exist and their effect on transport properties of dense matter should be considered as well.  The decay to three dark quarks \cite{Strumia:2021ybk,Husain:2022brl,Zhou:2023ndi} is particularly promising, as no repulsive self-interaction is needed.  Finally, quite apart from exotic neutron decays, the Urca rate with the NWA approximation has been calculated recently \cite{Alford:2024xfb}, and its effect on the Urca bulk viscosity should be examined.    
%%%%%%%%%%%%%%%%%%%%%%%%%%%%%%%%%%%%%%
\section*{Acknowledgments}
We would like to thank Gordon Baym, Glennys Farrar, Alex Haber, Cole Miller, Sanjay Reddy, and Brendan Reed for discussions, and Rana Nandi for providing the parameters used in the IUF-II EoS.  We also thank the Institute for Nuclear Theory at the University of Washington for its hospitality during the completion of this project.  CJH thanks the Aspen Center for Physics for its hospitality.  SPH acknowledges the support of the National Science Foundation grant PHY 21-16686 and CJH acknowledges the support of the US Department of Energy grant DE-FG02-87ER40365.
%%%%%%%%%%%%%%%%%%%%%%%%%%%%%%%%%
\appendix
%%%%%%%%%%%%%%%%%%%%%%%%%%%%%%
\begin{widetext}
\section{Neutron decay rate calculation}
%%%%%%%%%%%%%%%%%%%%%%%%%%%%%%%%%%
\subsection{Neutron direct dark decay rate} \label{sec:appendix_ndecay_direct}
The rate of $n\rightarrow\chi+\phi$ is
\begin{equation}
    \Gamma_{n\rightarrow\chi\phi}^{\text{direct}} = \int \dfrac{\mathop{d^3p_n}}{(2\pi)^3}\dfrac{\mathop{d^3p_{\chi}}}{(2\pi)^3}\dfrac{\mathop{d^3p_{\phi}}}{(2\pi)^3}(2\pi)^4\delta^4(p_n-p_{\chi}-p_{\phi})\dfrac{\sum_{\text{spins}}\vert\mathcal{M}\vert^2}{2^3E_n^*E_{\chi}^*E_{\phi}}f_n(1-f_{\chi})
\end{equation}
where the spin-summed matrix element is
\begin{equation}
    \sum_{\text{spins}}\vert\mathcal{M}\vert^2 = 4g_{\phi}^2\left(\tilde{p}_n\cdot\tilde{p}_{\chi}+m_*m_{\chi}\right).
\end{equation}
Integrating over $\mathbf{p_{\phi}}$ with the three-dimensional delta function gives
\begin{equation}
    \Gamma_{n\rightarrow\chi\phi}^{\text{direct}} = \dfrac{g_{\phi}^2}{64\pi^5}\int\mathop{d^3p_n}\mathop{d^3p_{\chi}}\delta\left[E_n-E_{\chi}-\sqrt{\left(\mathbf{p_n}-\mathbf{p_{\chi}}\right)^2+m_{\phi}^2}\right]\dfrac{\left(E_n^*E_{\chi}^*-\mathbf{p_n}\cdot\mathbf{p_{\chi}}+m_*m_{\chi}\right)f_n\left(1-f_{\chi}\right)}{E_n^*E_{\chi}^*\sqrt{\left(\mathbf{p_n}-\mathbf{p_{\chi}}\right)^2+m_{\phi}^2}}.
\end{equation}
We choose the z-axis to align with $\mathbf{p_n}$ and we choose to put $\mathbf{p_{\chi}}$ in the $xz$ plane
\begin{align}
    \mathbf{p_n} &= p_n\left(0,0,1\right)\\
    \mathbf{p_{\chi}} &= p_{\chi}\left(\sqrt{1-u^2},0,u\right),
\end{align}
where $u=\cos{\theta}$.  This choice of coordinates gives $8\pi^2$ from three trivial angular integrals.  We then integrate over the polar angle (cosine) $u$ using the energy delta function.  The delta function integral leads to the constraints
\begin{align}
    E_n&>E_{\chi}\label{eq:condition1}\\
\left\vert m_{\phi}^2-m_*^2-m_{\chi}^2-\Delta U^2-2\Delta U\left(E_n^*-E_{\chi}^*\right)+2E_n^*E_{\chi}^*\right\vert &\leq 2p_np_{\chi},    \label{eq:condition2}
\end{align}
and the integral becomes 
\begin{equation}
    \Gamma_{n\rightarrow\chi\phi}^{\text{direct}} = \dfrac{g_{\phi}^2}{16\pi^3}\int\mathop{dp_n}\mathop{dp_{\chi}}\dfrac{p_np_{\chi}}{E_n^*E_{\chi}^*}\left[ \left(m_*+m_{\chi}\right)^2-m_{\phi}^2+\Delta U^2+2\Delta U\left(E_n^*-E_{\chi}^*\right)  \right]f_n\left(1-f_{\chi}\right).
\end{equation}
This expression looks like it will simplify if one converts the momentum integrals to energy integrals, so
\begin{equation}
    \Gamma_{n\rightarrow\chi\phi}^{\text{direct}} = \dfrac{g_{\phi}^2}{16\pi^3}\int_{m_*}^{\infty}\mathop{dE_n^*}\int_{m_{\chi}}^{\infty}\mathop{dE_{\chi}^*}\left[ \left(m_*+m_{\chi}\right)^2-m_{\phi}^2+\Delta U^2+2\Delta U\left(E_n^*-E_{\chi}^*\right)  \right]f_n\left(1-f_{\chi}\right),
\end{equation}
subject to the constraints 
\begin{align}
    E_n^*+\Delta U&>E_{\chi}^*\label{eq:condition1new}\\
\left\vert m_{\phi}^2-m_*^2-m_{\chi}^2-\Delta U^2-2\Delta U\left(E_n^*-E_{\chi}^*\right)+2E_n^*E_{\chi}^*\right\vert &\leq 2\sqrt{(E_n^*)^2-m_*^2}\sqrt{(E_{\chi}^*)^2-m_{\chi}^2}.    \label{eq:condition2new}
\end{align}
%%%%%%%%%%%%%%%%%%%%%%%%%%%%%%%%%%
\subsection{Neutron ``modified'' dark decay rate: $n+n\rightarrow n+\chi+\phi$}\label{sec:appendix_ndecay_modified1}
The neutron can also decay to a $\chi$ and a $\phi$, but after first interacting with another neutron via the strong interaction.  There are two Feynman diagrams for this process, as the identical neutrons can be interchanged in the initial state.  In the calculation, the $n\chi\phi$ vertex is $ig_{\phi}$, and the strong interaction is modeled as a vector meson $\omega$ exchange, with the $nn\omega$ vertex $-ig_{\omega}\gamma^{\mu}$.  The $\omega$ meson is assumed to be heavy compared to the momentum transfer.  We define 
\begin{equation}
    %G_{\omega}\equiv\left(\dfrac{g_{\omega}}{m_{\omega}}\right)^2.
    G_{\omega}\equiv\left(g_{\omega}/m_{\omega}\right)^2.
\end{equation}
We also define the quantity
\begin{equation}
    \beta^2 \equiv (m_*+m_{\chi}+m_{\phi})(m_*+m_{\chi}-m_{\phi}).
\end{equation}
We will write the matrix element in terms of the 4-momentum transfers
\begin{align}
    k&\equiv p_1-p_3\\
    l&\equiv p_2-p_3.
\end{align}
All 4-momentum dot products can be written in terms of masses, $k$, $l$, and $p_{\phi}$
\begin{subequations}
\begin{align}
    p_1 \cdot p_2 &= m^2-\frac{1}{2}k^2-\frac{1}{2}l^2+k\cdot l\\
    p_1 \cdot p_3 &= m^2-\frac{1}{2}k^2\\
    p_1 \cdot p_4 &= \frac{1}{2}\left(m^2+m_{\chi}^2-m_{\phi}^2\right)-\frac{1}{2}l^2+l\cdot p_{\phi}\\
    p_2 \cdot p_3 &= m^2-\frac{1}{2}l^2\\
    p_2 \cdot p_4 &= \frac{1}{2}\left(m^2+m_{\chi}^2-m_{\phi}^2\right)-\frac{1}{2}k^2+k\cdot p_{\phi}\\
    p_3 \cdot p_4 &= \frac{1}{2}\left(m^2+m_{\chi}^2-m_{\phi}^2  \right)-\frac{1}{2}k^2-\frac{1}{2}l^2-k\cdot l +k\cdot p_{\phi}+l\cdot p_{\phi}\\
    p_1 \cdot p_{\phi} &= \frac{1}{2}\left( m^2-m_{\chi}^2+m_{\phi}^2 \right) + k\cdot l -l\cdot p_{\phi}\\
    p_2 \cdot p_{\phi} &= \frac{1}{2}\left( m^2-m_{\chi}^2+m_{\phi}^2 \right) +k\cdot l - k\cdot p_{\phi}\\
    p_3 \cdot p_{\phi} &= \frac{1}{2}\left( m^2-m_{\chi}^2+m_{\phi}^2 \right) +k\cdot l -k\cdot p_{\phi}-l\cdot p_{\phi}\\
    p_4\cdot p_{\phi} &= \frac{1}{2}\left( m^2-m_{\chi}^2-m_{\phi}^2 \right)+k\cdot l.
    \end{align}
\end{subequations}
These expressions are the generalization (to two different baryon masses $m$ and $m_{\chi}$) of those given in Eq.~A.4 in \cite{Dev:2020eam}.  With these approximations and definitions, the matrix element is given by
\begin{align}
    \sum_{\text{spins}}\vert\mathcal{M}\vert^2 &= \dfrac{2g_{\phi}^2G_{\omega}^2}{(k\cdot l)^2}\bigg\{ 8m_*^4\beta^2-8m_*^2\beta^2(k^2+l^2)+16m_*^3(m_*+m_{\chi})(k\cdot l)+5\beta^2(k^4+l^4)+8\beta^2k^2l^2\label{eq:matrix_element_4vecs_1}\\
    &-8m_*(m_*+m_{\chi})(k^2+l^2)(k\cdot l) +8(2m_*^2-m_*m_{\chi}-2m_{\chi}^2+2m_{\phi}^2)(k\cdot l)^2-16m_*^2(k\cdot l)\left[(k+l)\cdot p_{\phi}\right]\nonumber\\
    &-20(k^2+l^2)(k\cdot l)^2+32(k\cdot l)^3 +20(k\cdot l)\left[k^2(k\cdot p_{\phi})+l^2(l\cdot p_{\phi})\right]+16(k\cdot l)\left[k^2(l\cdot p_{\phi})+l^2(k\cdot p_{\phi})\right]\bigg\}.\nonumber
\end{align}
The symbols $k$, $l$, and $p_{\phi}$ are all 4-vectors in the above expression.  The neutron mass $m$ has been replaced with the nucleon effective mass $m_*$ because of the spin sum \cite{Roberts:2016mwj}.  In our model, the dark baryon $\chi$ does not self-interact in a scalar channel, so its mass remains just the vacuum value $m_{\chi}$.  In addition, we have neglected the vector mean fields in the nucleon propagator denominator, leading to the denominator being just $k\cdot l$.  

Next, we note that the energy transfer between nucleons, that is, $k_0$ and $l_0$, is quite small in degenerate matter.  So, we take $k^2\rightarrow -\mathbf{k}^2$, $l^2\rightarrow -\mathbf{l}^2$, $k\cdot l \rightarrow -\mathbf{k}\cdot\mathbf{l}$.  Also, terms like $k\cdot p_{\phi} = k_0E_{\phi}-\mathbf{k}\cdot\mathbf{p_{\phi}}\approx -\mathbf{k}\cdot\mathbf{p_{\phi}}$.  It is common to average over the direction of $p_{\phi}$, which causes all terms linear in $\mathbf{p_{\phi}}$ to vanish \cite{Brinkmann:1988vi}.  With these simplifications, the matrix element becomes
\begin{align}\label{eq:matrix_element_3vectors}
    \sum_{\text{spins}}\vert\mathcal{M}\vert^2 &= \dfrac{2g_{\phi}^2G_{\omega}^2}{(\mathbf{k}\cdot \mathbf{l})^2}\bigg\{ 8m_*^4\beta^2+8m_*^2\beta^2(\mathbf{k}^2+\mathbf{l}^2)-16m_*^3(m_*+m_{\chi})(\mathbf{k}\cdot \mathbf{l})+5\beta^2(\mathbf{k}^4+\mathbf{l}^4)+8\beta^2\mathbf{k}^2\mathbf{l}^2\\
    &-8m_*(m_*+m_{\chi})(\mathbf{k}^2+\mathbf{l}^2)(\mathbf{k}\cdot \mathbf{l}) +8(2m_*^2-m_*m_{\chi}-2m_{\chi}^2+2m_{\phi}^2)(\mathbf{k}\cdot \mathbf{l})^2+20(\mathbf{k}^2+\mathbf{l}^2)(\mathbf{k}\cdot \mathbf{l})^2-32(\mathbf{k}\cdot \mathbf{l})^3\bigg\}.\nonumber
\end{align}
The rate of the process $n+n\rightarrow n+\chi+\phi$ is given by the phase space integral
\begin{align}
    \Gamma = \int \dfrac{\mathop{d^3p_1}}{(2\pi)^3}\dfrac{\mathop{d^3p_2}}{(2\pi)^3}\dfrac{\mathop{d^3p_3}}{(2\pi)^3}\dfrac{\mathop{d^3p_4}}{(2\pi)^3}\dfrac{\mathop{d^3p_{\phi}}}{(2\pi)^3}\left(2\pi\right)^4\delta^4\left(p_1+p_2-p_3-p_4-p_{\phi}\right)\dfrac{S\sum_{\text{spins}}\vert\mathcal{M}\vert^2}{32E_1^*E_2^*E_3^*E_4^*E_{\phi}}f_1f_2(1-f_3)(1-f_4),\label{eq:ps_integral}
\end{align}
where $S=1/2$ is a symmetry factor to account for the identical neutron initial state, $E^*=E-U=\sqrt{p^2+m_*^2}$, and there is no Bose enhancement factor for the $\phi$ because it has a long mean free path in neutron star merger conditions, and thus escapes the system.  Having degenerate matter in mind, we multiply Eq.~\ref{eq:ps_integral} by unity in the form
\begin{align}
    1 &= \int_0^{\infty}\mathop{dp_1}\mathop{dp_2}\mathop{dp_3}\mathop{dp_4}\delta(p_1-p_{Fn})\delta(p_2-p_{Fn})\delta(p_3-p_{Fn})\delta(p_4-p_{F\chi})\\
    &= \frac{1}{p_{Fn}^3p_{F\chi}}\int\mathop{dE_1}\mathop{dE_2}\mathop{dE_3}\mathop{dE_4}E_1^*E_2^*E_3^*E_4^*\delta(p_1-p_{Fn})\delta(p_2-p_{Fn})\delta(p_3-p_{Fn})\delta(p_4-p_{F\chi}).
\end{align}
The $E_*$ factors here will cancel those under the matrix element in Eq.~\ref{eq:ps_integral}.  Eq.~\ref{eq:ps_integral} becomes
\begin{align}
    \Gamma &= \dfrac{1}{65536\pi^{11}}\dfrac{g_{\phi}^2G_{\omega}^2}{p_{Fn}^3p_{F\chi}}\int\mathop{d^3p_1}\mathop{d^3p_2}\mathop{d^3p_3}\mathop{d^3p_4}\mathop{d^3p_{\phi}}\mathop{dE_1}\mathop{dE_2}\mathop{dE_3}\mathop{dE_4}\delta(E_1+E_2-E_3-E_4-E_{\phi})\\
    &\times\delta^3(\mathbf{p_1}+\mathbf{p_2}-\mathbf{p_3}-\mathbf{p_4})\dfrac{1}{(\mathbf{k}\cdot\mathbf{l})^2}\dfrac{[...]}{E_{\phi}}\delta(p_1-p_{Fn})\delta(p_2-p_{Fn})\delta(p_3-p_{Fn})\delta(p_4-p_{F\chi})f_1f_2(1-f_3)(1-f_4),\nonumber
\end{align}
where $[...]$ represents the part of the matrix element inside the braces in Eq.~\ref{eq:matrix_element_3vectors}.  We also neglected $\mathbf{p_{\phi}}$ in the 3d delta function.  The utility of neglecting $\mathbf{p_{\phi}}$ in the 3d delta function and averaging over it in the matrix element now becomes clear.  We do the integral
\begin{equation}
    \int\mathop{d^3p_{\phi}} = 4\pi \int_0^{\infty}\mathop{dp_{\phi}}p_{\phi}^2\dfrac{1}{E_{\phi}}=4\pi\int_{m_{\phi}}^{\infty}\mathop{dE_{\phi}}\sqrt{E_{\phi}^2-m_{\phi}^2}.
\end{equation}
Now the rate integral can be split (``phase space decomposition'' \cite{Shapiro:1983du}) into two parts
\begin{equation}
    \Gamma = \dfrac{1}{16384\pi^{10}}\dfrac{g_{\phi}^2G_{\omega}^2}{p_{Fn}^3p_{F\chi}}AI,
\end{equation}
where $A$ and $I$ are an ``angular'' and an ``energy'' integral, defined below.  The energy integral, which will give rise to the temperature-dependence of the rate, is
\begin{equation}
    I = \int\mathop{dE_1}\mathop{dE_2}\mathop{dE_3}\mathop{dE_4}\mathop{dE_{\phi}}\delta(E_1+E_2-E_3-E_4-E_{\phi})\sqrt{E_{\phi}^2-m_{\phi}^2}f_1f_2(1-f_3)(1-f_4).
\end{equation}
We change variables, with $Tx_i \equiv E_i-\mu_i$ ($i$ from 1 to 4) and $Tx_{\phi}\equiv E_{\phi}$, leading to the integral
\begin{equation}
    I = T^5\int_{-(\mu_n-m)/T}^{\infty}\mathop{dx_1}\mathop{dx_2}\mathop{dx_3}\int_{-(\mu_{\chi}-m_{\chi})/T}^{\infty}\mathop{dx_4}\int_{m_{\phi}/T}^{\infty}\mathop{dx_{\phi}}\dfrac{\delta(x_1+x_2-x_3-x_4-x_{\phi}+\varepsilon)\sqrt{x_{\phi}^2-(m_{\phi}/T)^2}}{(1+e^{x_1})(1+e^{x_2})(1+e^{{-x_3}})(1+e^{-x_4})},
\end{equation}
where 
\begin{equation}
    \varepsilon \equiv \dfrac{\delta\mu}{T}\equiv \dfrac{\mu_n-\mu_{\chi}}{T}.
\end{equation}
In degenerate matter, $\mu-m\gg T$, and so the lower limit of the integrals over $x_1,x_2,x_3,x_4$ can be pushed down to $-\infty$.  Now, we do the integral
\begin{subequations}
\begin{align}
    I &= T^5\int_{-\infty}^{\infty}\mathop{dx_1}\mathop{dx_2}\mathop{dx_3}\int_{m_{\phi}/T}^{\infty}\mathop{dx_{\phi}}\dfrac{\sqrt{x_{\phi}^2-(m_{\phi}/T)^2}}{(1+e^{x_1})(1+e^{x_2})(1+e^{-x_3})(1+e^{x_3+x_{\phi}-x_1-x_2-\varepsilon})}\\
    &= T^5\int_{-\infty}^{\infty}\mathop{dx_1}\mathop{dx_2}\int_{m_{\phi}/T}^{\infty}\mathop{dx_{\phi}}\dfrac{\sqrt{x_{\phi}^2-(m_{\phi}/T)^2}\left(x_{\phi}-x_1-x_2-\varepsilon\right)}{(1+e^{x_1})(1+e^{x_2})(e^{x_{\phi}-x_1-x_2-\varepsilon}-1)}\\
    &= \dfrac{T^5}{2}\int_{-\infty}^{\infty}\mathop{dx_1}\int_{m_{\phi}/T}^{\infty}\mathop{dx_{\phi}}\dfrac{\sqrt{x_{\phi}^2-(m_{\phi}/T)^2}\left[\left(x_{\phi}-x_1-\varepsilon\right)^2+\pi^2\right]}{(1+e^{x_1})(1+e^{x_{\phi}-x_1-\varepsilon})}\\
    &=\dfrac{T^5}{6}\int_{m_{\phi}/T}^{\infty}\mathop{dx}\dfrac{\sqrt{x^2-(m_{\phi}/T)^2}\left(x-\varepsilon\right)\left[\left(x-\varepsilon\right)^2+4\pi^2\right]}{e^{x-\varepsilon}-1}.\label{eq:I_finite_mphi}
\end{align}
\end{subequations}
If $m_{\phi}=0$, this reduces to
\begin{equation}
    I\left(m_{\phi}=0\right) = \dfrac{T^5}{6}\left[-\varepsilon\left(\varepsilon^2+4\pi^2\right)\phi_2(e^{\varepsilon})+2\left(3\varepsilon^2+4\pi^2\right)\phi_3(e^{\varepsilon})-18\varepsilon\phi_4(e^{\varepsilon})+24\phi_5(e^{\varepsilon})\right],
\end{equation}
where $\phi_n$ is the Polylog of order $n$.  However, in this paper we keep $m_{\phi}$ finite and use Eq.~\ref{eq:I_finite_mphi}.

The angular integral is given by
\begin{equation}
    A = \int \mathop{d^3p_1}\mathop{d^3p_2}\mathop{d^3p_3}\mathop{d^3p_4}\delta^3\left(\mathbf{p_1}+\mathbf{p_2}-\mathbf{p_3}-\mathbf{p_4}\right)\dfrac{[...]}{\left(\mathbf{k}\cdot\mathbf{l}\right)^2}\delta(p_1-p_{Fn})\delta(p_2-p_{Fn})\delta(p_3-p_{Fn})\delta(p_4-p_{F\chi}),
\end{equation}
where $[...]$ again represents the part of the matrix element inside the braces in Eq.~\ref{eq:matrix_element_3vectors}, which is a function of masses, $\mathbf{k}^2$, $\mathbf{l}^2$, and $\mathbf{k}\cdot\mathbf{l}$.  Next, since the matrix element is made up of $\mathbf{k}$ and $\mathbf{l}$ combinations, we multiply by unity
\begin{equation}
    1 = \int\mathop{d^3k}\mathop{d^3l}\delta^3\left(\mathbf{k}-\mathbf{p_1}+\mathbf{p_3}\right)\delta^3\left(\mathbf{l}-\mathbf{p_2}+\mathbf{p_3}\right)
\end{equation}
and then integrate over $\mathbf{p_1}$ and $\mathbf{p_2}$, and then $\mathbf{p_4}$, leaving a nine-dimensional integral over $\mathbf{k}$, $\mathbf{l}$, and $\mathbf{p_3}$ (which we now just relabel as $\mathbf{p}$), leaving
\begin{align}
    A &= \int \mathop{d^3p}\mathop{d^3k}\mathop{d^3l}\frac{1}{\left(\mathbf{k}\cdot\mathbf{l}\right)^2}\delta\left(\vert\mathbf{p}+\mathbf{k}\vert-p_{Fn}\right)\delta\left(\vert\mathbf{p}+\mathbf{l}\vert-p_{Fn}\right)\delta\left(p-p_{Fn}\right)\delta\left(\vert\mathbf{p}+\mathbf{k}+\mathbf{l}\vert-p_{F\chi}\right)\\
    &\bigg\{ 8m_*^4\beta^2+8m_*^2\beta^2(\mathbf{k}^2+\mathbf{l}^2)-16m_*^3(m_*+m_{\chi})(\mathbf{k}\cdot \mathbf{l})+5\beta^2(\mathbf{k}^4+\mathbf{l}^4)+8\beta^2\mathbf{k}^2\mathbf{l}^2\nonumber\\
    &-8m_*(m_*+m_{\chi})(\mathbf{k}^2+\mathbf{l}^2)(\mathbf{k}\cdot \mathbf{l}) +8(2m_*^2-m_*m_{\chi}-2m_{\chi}^2+2m_{\phi}^2)(\mathbf{k}\cdot \mathbf{l})^2+20(\mathbf{k}^2+\mathbf{l}^2)(\mathbf{k}\cdot \mathbf{l})^2-32(\mathbf{k}\cdot \mathbf{l})^3\bigg\}\nonumber.
\end{align}
We have the freedom to align $\mathbf{p}$ along the z axis, and so we can use the coordinate system
\begin{align}
    \mathbf{p}&=p(0,0,1)\\
    \mathbf{k}&=k(\sqrt{1-r^2},0,r)\nonumber\\
    \mathbf{l}&=l(\sqrt{1-s^2}\cos{\phi},\sqrt{1-s^2}\sin{\phi},s),\nonumber
\end{align}
where $-1\leq r,s\leq 1$ and $0\leq\phi<2\pi$.

Next, we do the integrals over $r$ and $s$, eliminating two of the delta functions, and picking up constraints $k<2p_{Fn}$ and $l<2p_{Fn}$.  We change variables to $a \equiv k/(2p_{Fn})$ and $b \equiv l/(2p_{Fn})$ and also define
\begin{equation}
    \alpha \equiv p_{F\chi}/p_{Fn}.
\end{equation}
We next encounter the integral over $\phi$, which has the form
\begin{equation}
I_{\phi} =\int_0^{2\pi}\mathop{d\phi}\delta\left(\sqrt{1+8a^2b^2+8ab\sqrt{1-a^2}\sqrt{1-b^2}\cos{\phi}}-\alpha\right)f(\phi),
\end{equation}
where $f(\phi)$ represents the complicated matrix element expression to the extent that it depends on $\phi$.  This delta function has two zeros in the integration interval, provided 
\begin{equation}
    \vert \alpha^2-1-8a^2b^2\vert \leq 8ab\sqrt{1-a^2}\sqrt{1-b^2} \label{eq:appendix_constraint}
\end{equation}
The integral yields 
\begin{equation}
    I_{\phi} = \dfrac{4\alpha f(\phi_0)}{\sqrt{64a^2b^2(1-a^2)(1-b^2)-(\alpha^2-1-8a^2b^2)^2}}.
\end{equation}
Now we are prepared to write down the expression for the angular integral $A$ including the complicated matrix element expression.  Let us note that now, after various integrations and coordinate transformations, $\mathbf{k}^2 = 4p_{Fn}^2a^2$, $\mathbf{l}^2=4p_{Fn}^2b^2$, and $\mathbf{k}\cdot\mathbf{l}=p_{Fn}^2(\alpha^2-1)/2.$  Now, 
\begin{align}
    A &= \dfrac{4096\pi^2p_{F\chi}}{(\alpha^2-1)^2}\int_0^1\mathop{da}\mathop{db}ab\bigg[4m_*^4\beta^2+16m_*^2\beta^2p_{Fn}^2(a^2+b^2)-4m_*^3(m_*+m_{\chi})p_{Fn}^2(\alpha^2-1)+40\beta^2p_{Fn}^4(a^4+b^4)\nonumber\\
    &+64\beta^2p_{Fn}^4a^2b^2-8m_*(m_*+m_{\chi})p_{Fn}^4(a^2+b^2)(\alpha^2-1)+\left(2m_*^2-m_*m_{\chi}-2m_{\chi}^2+2m_{\phi}^2\right)p_{Fn}^4(\alpha^2-1)^2\nonumber\\
    &+10p_{Fn}^6(a^2+b^2)(\alpha^2-1)^2-2p_{Fn}^6(\alpha^2-1)^3\bigg]/\sqrt{64a^2b^2(1-a^2)(1-b^2)-(\alpha^2-1-8a^2b^2)^2}.
\end{align}
subject to the constraint Eq.~\ref{eq:appendix_constraint}.  Finally, we switch variables to $x=a^2$ and $y=b^2$, obtaining
\begin{align}
    A &= \dfrac{1024\pi^2p_{F\chi}}{(\alpha^2-1)^2}\int_0^1\mathop{dx}\mathop{dy}\bigg[4m_*^4\beta^2+16m_*^2\beta^2p_{Fn}^2(x+y)-4m_*^3(m_*+m_{\chi})p_{Fn}^2(\alpha^2-1)+40\beta^2p_{Fn}^4(x^2+y^2)\nonumber\\
    &+64\beta^2p_{Fn}^4xy-8m_*(m_*+m_{\chi})p_{Fn}^4(\alpha^2-1)(x+y)+\left(2m_*^2-m_*m_{\chi}-2m_{\chi}^2+2m_{\phi}^2\right)p_{Fn}^4(\alpha^2-1)^2\nonumber\\
    &+10p_{Fn}^6(\alpha^2-1)^2(x+y)-2p_{Fn}^6(\alpha^2-1)^3\bigg]/\sqrt{64xy(1-x)(1-y)-(\alpha^2-1-8xy)^2}
\end{align}
subject to
\begin{equation}
    \vert \alpha^2-1-8xy\vert \leq 8\sqrt{xy}\sqrt{1-x}\sqrt{1-y}.\label{eq:appendix_constraint_xy}
\end{equation}
Doing the integration, we find that the answer splits into two cases: $0<\alpha\leq 1$ and $1<\alpha\leq 3$.  If $\alpha>3$, the rate is zero in degenerate nuclear matter.  The case $0<\alpha<1$ is the one relevant for the physical situation in this paper.  We find
\begin{equation}
    A = 1024\pi^3p_{F\chi}\times\label{eq:nnnchiphi_angular_integral}
    \begin{cases}
        g_1(m_*,m_{\chi},m_{\phi},p_{Fn},p_{F\chi}), & 0\leq \alpha < 1\\
        g_2(m_*,m_{\chi},m_{\phi},p_{Fn},p_{F\chi}), & 1<\alpha < 3\\
        0, & \alpha >3
    \end{cases}
\end{equation}   
where
\begin{align}
g_1(m_*,m_{\chi},m_{\phi},p_{Fn},p_{F\chi}) &= \dfrac{\alpha}{\left(1-\alpha^2\right)^2} \bigg[m_*^4\beta^2+\dfrac{2}{3}m_*^2\beta^2p_{Fn}^2\left(3+\alpha^2\right)+m_*^3\left(m_*+m_{\chi}\right)p_{Fn}^2\left(1-\alpha^2\right)\\
&+\dfrac{1}{3}m_*\left(m_*+m_{\chi}\right)p_{Fn}^4\left(1-\alpha^2\right)\left(3+\alpha^2\right)+\dfrac{1}{60}\beta^2p_{Fn}^4\left(135+150\alpha^2+19\alpha^4\right)\nonumber\\
&+\dfrac{1}{4}\left(2m_*-m_*m_{\chi}-2m_{\chi}^2+2m_{\phi}^2\right)p_{Fn}^4\left(1-\alpha^2\right)^2+\dfrac{1}{12}p_{Fn}^6\left(21-\alpha^2\right)\left(1-\alpha^2\right)^2\bigg]\nonumber\\
g_2(m_*,m_{\chi},m_{\phi},p_{Fn},p_{F\chi}) &= \dfrac{3-\alpha}{\left(\alpha^2-1\right)^2}\bigg[\dfrac{1}{2}m_*^4\beta^2+\dfrac{1}{3}m_*^2\beta^2p_{Fn}^2\left(3+\alpha^2\right)-\dfrac{1}{2}m_*^3\left(m_*+m_{\chi}\right)p_{Fn}^2\left(\alpha^2-1\right)\\
&-\dfrac{1}{6}m_*\left(m_*+m_{\chi}\right)p_{Fn}^4\left(\alpha^2-1\right)\left(3+\alpha^2\right)+\dfrac{1}{120}\beta^2p_{Fn}^4\left(189+18\alpha+96\alpha^2-18\alpha^3+19\alpha^4\right)\nonumber\\
&+\dfrac{1}{8}\left(2m_*^2-m_*m_{\chi}-2m_{\chi}^2+2m_{\phi}^2\right)p_{Fn}^4\left(\alpha^2-1\right)^2+\dfrac{1}{24}p_{Fn}^6\left(21-\alpha^2\right)\left(\alpha^2-1\right)^2\bigg].\nonumber
\end{align}
Thus, we get for the final rate, valid for arbitrary $\delta\mu$ departures from chemical equilibrium,
\begin{equation}
    \Gamma = \dfrac{1}{96\pi^7}\dfrac{g_{\phi}^2G_{\omega}^2}{p_{Fn}^3}T^5J(m_{\phi},T,\delta\mu)\times \begin{cases}
        g_1(m_*,m_{\chi},m_{\phi},p_{Fn},p_{F\chi}), & 0\leq \alpha < 1\\
        g_2(m_*,m_{\chi},m_{\phi},p_{Fn},p_{F\chi}), & 1<\alpha < 3\\
        0, & \alpha >3
    \end{cases}
\end{equation}
where 
\begin{equation}
    J(m_{\phi},T,\delta\mu) = \int_{m_{\phi}/T}^{\infty}\mathop{dx}\dfrac{\sqrt{x^2-(m_{\phi}/T)^2}\left(x-\varepsilon\right)\left[\left(x-\varepsilon\right)^2+4\pi^2\right]}{e^{x-\varepsilon}-1}
\end{equation}
%%%%%%%%%%%%%%%%%%%%%%%%%%%%%%%%%%%%%%%
\subsection{Neutron ``modified'' dark decay rate: $n+\chi\rightarrow \chi+\chi+\phi$}\label{sec:appendix_ndecay_modified2}
The neutron can decay to a $\chi$ and a $\phi$, but where the $\chi$ is off-shell, and then scatters with another $\chi$ to bring it back on-shell.  There are two Feynman diagrams for this process, as the identical dark baryons $\chi$ can be interchanged in the final state.  The dark baryons $\chi$ are assumed to interact repulsively by exchanging a dark vector boson, the $\omega'$.  Recall the definition of the dark baryon self-interaction strength $G'$, given in Eq.~\ref{eq:Gprime}.

Defining $k$ and $l$ in the same way as above, we find
\begin{subequations}
\begin{align}
    p_1 \cdot p_2 &= m_{\chi}^2-\frac{1}{2}k^2-\frac{1}{2}l^2+k\cdot l\\
    p_1 \cdot p_3 &= m_{\chi}^2-\frac{1}{2}k^2\\
    p_1 \cdot p_4 &= \frac{1}{2}\left(m^2+m_{\chi}^2-m_{\phi}^2\right)-\frac{1}{2}l^2-l\cdot p_{\phi}\\
    p_2 \cdot p_3 &= m_{\chi}^2-\frac{1}{2}l^2\\
    p_2 \cdot p_4 &= \frac{1}{2}\left(m^2+m_{\chi}^2-m_{\phi}^2\right)-\frac{1}{2}k^2-k\cdot p_{\phi}\\
    p_3 \cdot p_4 &= \frac{1}{2}\left(m^2+m_{\chi}^2-m_{\phi}^2  \right)-\frac{1}{2}k^2-\frac{1}{2}l^2-k\cdot l -k\cdot p_{\phi}-l\cdot p_{\phi}\\
    p_1 \cdot p_{\phi} &= \frac{1}{2}\left( m^2-m_{\chi}^2-m_{\phi}^2 \right) - k\cdot l -l\cdot p_{\phi}\\
    p_2 \cdot p_{\phi} &= \frac{1}{2}\left( m^2-m_{\chi}^2-m_{\phi}^2 \right) -k\cdot l - k\cdot p_{\phi}\\
    p_3 \cdot p_{\phi} &= \frac{1}{2}\left( m^2-m_{\chi}^2-m_{\phi}^2 \right) -k\cdot l -k\cdot p_{\phi}-l\cdot p_{\phi}\\
    p_4\cdot p_{\phi} &= \frac{1}{2}\left( m^2-m_{\chi}^2+m_{\phi}^2 \right)-k\cdot l.
    \end{align}
\end{subequations}
The spin-summed matrix element is given by
\begin{align}
    \sum_{\text{spins}}\vert\mathcal{M}\vert^2 &= \dfrac{2g_{\phi}^2G'^2}{(k\cdot l)^2}\bigg\{ 8m_{\chi}^4\beta^2-8m_{\chi}^2\beta^2(k^2+l^2)+16m_{\chi}^3(m_*+m_{\chi})(k\cdot l)+5\beta^2(k^4+l^4)+8\beta^2k^2l^2\\
    &-8m_{\chi}(m_*+m_{\chi})(k^2+l^2)(k\cdot l) +8(2m_{\chi}^2-m_*m_{\chi}-2m_*^2+2m_{\phi}^2)(k\cdot l)^2+16m_{\chi}^2(k\cdot l)\left[(k+l)\cdot p_{\phi}\right]\nonumber\\
    &-20(k^2+l^2)(k\cdot l)^2+32(k\cdot l)^3 -20(k\cdot l)\left[k^2(k\cdot p_{\phi})+l^2(l\cdot p_{\phi})\right]-16(k\cdot l)\left[k^2(l\cdot p_{\phi})+l^2(k\cdot p_{\phi})\right]\bigg\}.\nonumber
\end{align}
Notice that this matrix element is the same as Eq.~\ref{eq:matrix_element_4vecs_1} except with $m_*$ and $m_{\chi}$ interchanged, with $p_{\phi}\rightarrow - p_{\phi}$, and $G'$ replacing $G_{\omega}$.  Now we do the same as before: neglect the zeroth components of the 4-vectors $k$ and $l$, and neglect terms with $p_{\phi}$, obtaining
\begin{align}\label{eq:matrix_element_3vectors_2}
    \sum_{\text{spins}}\vert\mathcal{M}\vert^2 &= \dfrac{2g_{\phi}^2G'^2}{(\mathbf{k}\cdot \mathbf{l})^2}\bigg\{ 8m_{\chi}^4\beta^2+8m_{\chi}^2\beta^2(\mathbf{k}^2+\mathbf{l}^2)-16m_{\chi}^3(m_*+m_{\chi})(\mathbf{k}\cdot \mathbf{l})+5\beta^2(\mathbf{k}^4+\mathbf{l}^4)+8\beta^2\mathbf{k}^2\mathbf{l}^2\\
    &-8m_{\chi}(m_*+m_{\chi})(\mathbf{k}^2+\mathbf{l}^2)(\mathbf{k}\cdot \mathbf{l}) +8(2m_{\chi}^2-m_*m_{\chi}-2m_*^2+2m_{\phi}^2)(\mathbf{k}\cdot \mathbf{l})^2+20(\mathbf{k}^2+\mathbf{l}^2)(\mathbf{k}\cdot \mathbf{l})^2-32(\mathbf{k}\cdot \mathbf{l})^3\bigg\}.\nonumber
\end{align}
This expression is the same as Eq.~\ref{eq:matrix_element_3vectors} but with $m_*$ and $m_{\chi}$ interchanged and $G'$ replacing $G_{\omega}$.

The rate of the process $n+\chi\rightarrow \chi+\chi+\phi$ is given by the phase space integral
\begin{align}
    \Gamma = \int \dfrac{\mathop{d^3p_1}}{(2\pi)^3}\dfrac{\mathop{d^3p_2}}{(2\pi)^3}\dfrac{\mathop{d^3p_3}}{(2\pi)^3}\dfrac{\mathop{d^3p_4}}{(2\pi)^3}\dfrac{\mathop{d^3p_{\phi}}}{(2\pi)^3}\left(2\pi\right)^4\delta^4\left(p_1+p_2-p_3-p_4+p_{\phi}\right)\dfrac{S\sum_{\text{spins}}\vert\mathcal{M}\vert^2}{32E_1^*E_2^*E_3^*E_4^*E_{\phi}}(1-f_1)(1-f_2)f_3f_4,\label{eq:ps_integral_2}.
\end{align}
Following the same steps as in the $n+n\rightarrow n + \chi+\phi$ case, we find
\begin{equation}
    \Gamma = \dfrac{1}{16384\pi^{10}}\dfrac{g_{\phi}^2G'^2}{p_{Fn}p_{F\chi}^3}AI,
\end{equation}
where $A$ and $I$ are defined by 
\begin{equation}
    I = \int\mathop{dE_1}\mathop{dE_2}\mathop{dE_3}\mathop{dE_4}\mathop{dE_{\phi}}\delta(E_1+E_2-E_3-E_4+E_{\phi})\sqrt{E_{\phi}^2-m_{\phi}^2}(1-f_1)(1-f_2)f_3f_4
\end{equation}
and 
\begin{equation}
    A = \int \mathop{d^3p_1}\mathop{d^3p_2}\mathop{d^3p_3}\mathop{d^3p_4}\delta^3\left(\mathbf{p_1}+\mathbf{p_2}-\mathbf{p_3}-\mathbf{p_4}\right)\dfrac{[...]}{\left(\mathbf{k}\cdot\mathbf{l}\right)^2}\delta(p_1-p_{F\chi})\delta(p_2-p_{F\chi})\delta(p_3-p_{F\chi})\delta(p_4-p_{Fn}),
\end{equation}
where $[...]$ again represents the part of the matrix element inside the braces in Eq.~\ref{eq:matrix_element_3vectors_2}.  We find
\begin{equation}
    I =\dfrac{T^5}{6}\int_{m_{\phi}/T}^{\infty}\mathop{dx}\dfrac{\sqrt{x^2-(m_{\phi}/T)^2}\left(x-\varepsilon\right)\left[\left(x-\varepsilon\right)^2+4\pi^2\right]}{e^{x-\varepsilon}-1},\label{eq:I_finite_mphi_2}
\end{equation}
the same expression as in the $n+n\rightarrow n+\chi+\phi$ case.  In evaluating the angular integral, it is convenient to define
\begin{equation}
    \tilde{\alpha}\equiv \alpha^{-1} = p_{Fn}/p_{F\chi}.
\end{equation}
We find
\begin{equation}
    A = 1024\pi^3p_{Fn}\times
    \begin{cases}
        g_3(m_*,m_{\chi},m_{\phi},p_{Fn},p_{F\chi}), & 0\leq \tilde{\alpha} < 1\\
        g_4(m_*,m_{\chi},m_{\phi},p_{Fn},p_{F\chi}), & 1<\tilde{\alpha} < 3\\
        0, & \tilde{\alpha} >3,
    \end{cases}
\end{equation}   
where
\begin{align}
g_3(m_*,m_{\chi},m_{\phi},p_{Fn},p_{F\chi}) &= \dfrac{\tilde{\alpha}}{\left(1-\tilde{\alpha}^2\right)^2} \bigg[m_{\chi}^4\beta^2+\dfrac{2}{3}m_{\chi}^2\beta^2p_{F\chi}^2\left(3+\tilde{\alpha}^2\right)+m_{\chi}^3\left(m_*+m_{\chi}\right)p_{F\chi}^2\left(1-\tilde{\alpha}^2\right)\\
&+\dfrac{1}{3}m_{\chi}\left(m_*+m_{\chi}\right)p_{F\chi}^4\left(1-\tilde{\alpha}^2\right)\left(3+\tilde{\alpha}^2\right)+\dfrac{1}{60}\beta^2p_{F\chi}^4\left(135+150\tilde{\alpha}^2+19\tilde{\alpha}^4\right)\nonumber\\
&+\dfrac{1}{4}\left(2m_{\chi}^2-m_*m_{\chi}-2m_*^2+2m_{\phi}^2\right)p_{F\chi}^4\left(1-\tilde{\alpha}^2\right)^2+\dfrac{1}{12}p_{F\chi}^6\left(21-\tilde{\alpha}^2\right)\left(1-\tilde{\alpha}^2\right)^2\bigg]\nonumber\\
g_4(m_*,m_{\chi},m_{\phi},p_{Fn},p_{F\chi}) &= \dfrac{3-\tilde{\alpha}}{\left(\tilde{\alpha}^2-1\right)^2}\bigg[\dfrac{1}{2}m_{\chi}^4\beta^2+\dfrac{1}{3}m_{\chi}^2\beta^2p_{F\chi}^2\left(3+\tilde{\alpha}^2\right)-\dfrac{1}{2}m_{\chi}^3\left(m_*+m_{\chi}\right)p_{F\chi}^2\left(\tilde{\alpha}^2-1\right)\\
&-\dfrac{1}{6}m_{\chi}\left(m_*+m_{\chi}\right)p_{F\chi}^4\left(\tilde{\alpha}^2-1\right)\left(3+\tilde{\alpha}^2\right)+\dfrac{1}{120}\beta^2p_{F\chi}^4\left(189+18\tilde{\alpha}+96\tilde{\alpha}^2-18\tilde{\alpha}^3+19\tilde{\alpha}^4\right)\nonumber\\
&+\dfrac{1}{8}\left(2m_{\chi}^2-m_*m_{\chi}-2m_*^2+2m_{\phi}^2\right)p_{F\chi}^4\left(\tilde{\alpha}^2-1\right)^2+\dfrac{1}{24}p_{F\chi}^6\left(21-\tilde{\alpha}^2\right)\left(\tilde{\alpha}^2-1\right)^2\bigg]\nonumber,
\end{align}
which is nothing but the angular integral Eq.~\ref{eq:nnnchiphi_angular_integral} but with $p_{Fn}$ and $p_{F\chi}$ interchanged, as well as $m_*$ and $m_{\chi}$ interchanged.  The rate of $n+\chi \rightarrow \chi+\chi+\phi$ is
\begin{equation}
    \Gamma = \dfrac{1}{96\pi^7}\dfrac{g_{\phi}^2G'^2}{p_{F\chi}^3}T^5J(m_{\phi},T,\delta\mu)\times \begin{cases}
        g_3(m_*,m_{\chi},m_{\phi},p_{Fn},p_{F\chi}), & 0\leq \tilde{\alpha} < 1\\
        g_4(m_*,m_{\chi},m_{\phi},p_{Fn},p_{F\chi}), & 1<\tilde{\alpha} < 3\\
        0, & \tilde{\alpha} >3
    \end{cases}.
\end{equation}
\end{widetext}
\bibliography{references}
\end{document}